\documentclass[pdftex,twocolumn,epjc3]{svjour3}

\input{preamble}

\journalname{International Journal of Information Security}

\begin{document}

\title{Man-at-the-End Software Protection as a Risk Analysis Process}

\author{Daniele Canavese\thanksref{e1,addr1}
  \and Leonardo Regano\thanksref{e1,addr1}
  \and Cataldo Basile\thanksref{e1,addr1}
  \and \\ Bart Coppens\thanksref{e2,addr2}
  \and Bjorn De Sutter\thanksref{e2,addr2}
}

\thankstext{e1}{e-mail:\{daniele.cavanese,leonardo.regano,cataldo.basile\}@polito.it}
\thankstext{e2}{e-mail:\{bart.coppens,bjorn.desutter\}@ugent.be}

\institute{  Dipartimento di Automatica e Informatica, Politecnico di Torino, Corso Duca degli Abruzzi, 24, 10129 Torino, Italy\label{addr1}
  \and
  Computer Systems Lab, Ghent University, Technologiepark-Zwijnaarde 126, 9052 Zwijnaarde, Belgium\label{addr2}
}

%\author{Bart Coppens}
%\author{Bjorn De Sutter}
%\email{firstname.lastname@ugent.be}
%\affiliation{%
%	\institution{Computer Systems Lab, Ghent University}
%	\streetaddress{Technologiepark Zwijnaarde 126}
%	\city{Ghent}
%	\country{Belgium}
%	\postcode{9052}
%}


\maketitle

\begin{abstract}
  The last years have seen an increase of Man-at-the-End (MATE) attacks against software applications, both in number and severity.
  However, MATE software protections are dominated by fuzzy concepts and techniques, with security-through-obscurity omnipresent in the field.
  This paper presents a rationale for adopting and standardizing the protection of software as a risk management process according to the NIST SP800-39 approach.
  We examine the relevant aspects of formalizing and automating the activities in this process in the context of MATE software protection. We highlight the open issues that the research community still has to address. We discuss the benefits that such an approach can bring to all stakeholders.
  In addition, we present a Proof of Concept (PoC) of a decision support system that automates many activities in the risk analysis methodology towards the protection of software applications.
  {Despite still being a prototype, the PoC validation with industry experts indicated that several aspects of the proposed risk management process can already be formalized and automated with our existing toolbox, and that it can actually assist decision making in industrially relevant settings.}
\end{abstract}




\section{Introduction}
\label{sec:intro}
{In the \mate attack model, attackers have white-box access to software, meaning that they have full control over the systems on which they attack the software and on which they can hence run all kinds of attacker tools, such as simulators, debuggers, disassemblers, decompilers, and other kinds of static and dynamic analysis and tampering tools. Attackers can use those tools to inspect and alter stored software as well as the internal state of running software. They perform such actions to reverse engineer the software (e.g., to steal valuable secret algorithms or embedded cryptographic keys, or to find vulnerabilities in the code), that they try to tamper with the software (e.g., to bypass license checks or to cheat in games), or that try to execute them in unauthorized ways (e.g., run multiple copies in parallel). In general, in the \mate attack model, attacks target software components to violate \emph{security requirements of assets present in those software components themselves}.

In this context, {\mate} {\softprot}, often shortened to {\softprot} in the remainder of this paper, refers to protections deployed \emph{within those software components} to mitigate such \mate attacks. {\mate} {\softprot} is hence much narrower than the broad umbrella of software security. The latter also includes scenarios in which software components are exploited to violate \emph{security requirements of other system components}, e.g., to infiltrate networks or to escalate privileges. In such scenarios, attackers start with much more limited capabilities (e.g., only unprivileged, remote access via a web server interface). Those attackers are then often not the owners of the software being exploited or of the devices and networks on which the exploited software runs. 

{\mate} {\softprot} to the contrary needs to defend assets in the software against attackers with full access and privileged control over the devices on which they engineer their attacks, i.e., as they identify successful attack vectors in their lab. As such, {\mate} {\softprot} \emph{cannot rely on external services} running on those devices. Instead only protections deployed within the protected software itself or on remote servers controlled by the defender can be relied upon. 
}

Advances in cryptography have yielded techniques such as multilinear jigsaw puzzles~\cite{JigSawObfuscation2013} that provide strong security guarantees against reverse engineering based on concepts like homomorphic encryption~\cite{GentryPhD}, but those introduce orders of magnitude performance overhead~\cite{fasterpackedhomomorphic2017}. 
Hence, in practice, they are rarely usable today. 
Instead, {\softprot} is dominated by fuzzy concepts and techniques~\cite{collbergbook}. {\softprot}s such as remote attestation, obfuscation, anti-debugging, software diversity, and anti-emulation do not aim to mitigate \mate attacks completely.
Instead they aim to delay attacks and put off potential attackers by making attacks expensive enough and the expected \roi low enough. As observed during a recent Dagstuhl seminar on \softprot Decision Support and Evaluation Methodologies~\cite{Dagstuhl}, the \softprot field is facing severe challenges: security-through-obscurity is omnipresent in industry; 
{\softprot} tools and consultancy are expensive and opaque; 
there is no generally accepted method for evaluating {\softprot}s and {\softprot} tools. Moreover, {\softprot} tools are not deployed sufficiently~\cite{arxan-report,ceccato-new-one,GOP,BSA}; expertise is largely missing in software vendors to deploy (third-party) {\softprot} tools~\cite{Gartner-report-online,Irdeto-report1,Mandiant}; 
and we lack standardization. 
The \nist SP800-39 IT systems risk management standard~\cite{nistSP800-39} or the ISO27k framework for information risk management~\cite{ISO27k}, which are deployed consistently in practice to secure corporate computer networks, have no counterpart or instance in the field of {\softprot}. 
Neither do we have concrete regulations to implement \gdpr compliance in applications.% sensitive to \mate attacks. 

The goal of our research is to explore how a standardized risk management approach can benefit the domain of \softprot, and to identify already available technology as well as open problems on which further research is needed. Moreover, we want to address the issue of automating the application of the {\softprot} risk analysis. As software developers do not have the resources to manage a risk analysis because of a lack of in-house competences in {\softprot} techniques and because there are no regulations forcing them, an automatic process could reduce the required effort and expertise, and make {\softprot} accessible.
%\abnote{not sure about the objective, the first objective is already the result of the assumption that standardization is a solution. I would rephrase the objectives: we present why a standardized risk management is useful and show that it is in the end feasible. Need to discuss with you though.}
%With \changed{our vision and observations as reported in} this paper, we hope to instigate the necessary actions towards a standardized risk management approach. % for \softprot that follows the principle of the \nist 800 series. 
%Furthermore, we address the issue of automating the application of the {\softprot} risk analysis. As software developers do not have the resources to manage a risk analysis because of a lack of in-house competences in {\softprot} techniques and because there are no regulations forcing them, an automatic process could reduce the required effort and expertise, and make {\softprot} accessible.
%and have a positive effect on everyone's security, as properly protected software can reduce the likelihood of attacks \abnote{Cicero warning}. 

Towards these goals, this paper offers a number of contributions.
First, we provide a rationale for adopting and standardizing risk management processes for {\softprot}. We discuss a number of observations on the failing {\softprot} market and analyse why the existing standards as adopted in network security are not applicable for {\softprot}, i.e., what makes {\softprot} different from network security from the perspective of risk management.

Secondly, we discuss in depth how to adopt the \nist risk management approach. For all the required risk management processes, we highlight (i) the current status; (ii) the \softprot-specific concepts and aspects that need to be covered; (iii) what existing parts can be borrowed from other fields; (iv) where we see open questions and challenges that require further research; (v) where we see the need for the research community and industry to come together to define standards; and (vi) the relevant aspect towards formalizing and automating the discussed risk management activities. 

Last but not least, 
%and despite not yet having a complete picture or a complete instance or standardization of the whole proposed methodology, 
we demonstrate that several aspects can already be formalized and automated. To that extent, we present a \poc decision support system that automates some of the major risk management activities and processes. 
This tool is able to drive the identification, assessment, and mitigation of risks. Even if not completely automated, it provides a starting point for protecting applications and for building a more advanced system that follows all the methodological aspects of a \nist 800-compliant standard and reaches industrial grade maturity. The first results obtained with the tool have been validated by industry experts on a number of Android mobile app case studies of real-world complexity. 

The remainder of the paper is structured along those three contributions.

\section{Rationale for Proper Risk Management}
\label{sec:rationale}
We first discuss some risk management standards from other security domains such as network security, and the healthy market for products and services that exists there as a result. We then contrast this with the lack of such a market and standards for {\softprot}, for which we discuss the challenges to make progress towards proper risk management standards. Finally, we highlight what benefits such progress can bring.

\subsection{Standardized risk management approaches}
\label{sec:standardizedRisk}
Protecting software may be more formally defined by framing it as a risk management process, a customary activity in various industries such as finance, pharmaceutics, infrastructure, energy and \itech. 
Regarding the latter, the \nist has proposed an \itech systems risk management standard that sets the context and identifies four main phases \cite{nistSP800-39}:
%\begin{enumerate}
    (i) \emph{risk framing}: to establish the scenario in which the risk must be managed;
	(ii) \emph{risk assessment}: to identify threats against the system assets, vulnerabilities of the system, the harm that may occur if those are exploited, and the likelihood thereof;
	(iii)  \emph{risk mitigation}: to determine and implement appropriate actions to mitigate the risks;
	{and}
	(iv) \emph{risk monitoring}: to verify that the implemented actions are effective in mitigating the risks.
%\end{enumerate}

The ISO27k framework also focuses on information risk management, to be managed in three phases~\cite{ISO27k}:
(i) \emph{identify risk} to identify the main threats and vulnerabilities that loom over assets; (ii) 
\emph{evaluate risk} to estimate the impact of the consequences of the risks; and (iii)
\emph{treat risk} to mitigate the risks that can be neither accepted nor avoided.
The ISO27k framework adds an explicit operational phase for handling changes that may happen in the framed risk scenario.

Those approaches have been consistently applied in practice for securing corporate networks.
Regulations stimulated companies to analyse the risks against their \itech systems. For instance, the \gdpr explicitly requires a risk analysis of all private data handling. 
Companies invest in compliance with the ISO27k family, as it provides market access.
As a consequence, risk analysis of networks has developed a common vocabulary and a company's tasks have been properly identified and often standardized, so offerings from consultancy firms can be compared easily. 
There is a business related to this task, best practices, and big consultant firms have risk analysis of corporate networks in their catalogs~\cite{Gartner-report-riskanalysis}.

In the domain of software security, several frameworks for risk analysis and decision support exist, which mainly focus on Software Vulnerability Management~\cite{nistir8011} and Enterprise Patch Management~\cite{nistSP800-40}. 
Moreover, other frameworks focus on software quality assurance best practices and benchmarking. including the OWASP Software Assurance Maturity Model (SAMM)~\cite{owaspsamm}, the OWASP Application Security Verification Standard (ASVS)~\cite{owaspasvs}, and the Building Security in Maturity Model (BSIMM)~\cite{bsimm}. 
These frameworks address problems of software security and are not applicable to {\softprot}.

NIST SP800-53~\cite{nistSP800-53} extends beyond software security and provides a comprehensive and flexible catalog of privacy and security controls for systems and organizations as part of their organizational risk mitigation strategy, for which they build on NIST SP800-39~\cite{nistSP800-39}. It targets their whole IT infrastructure, including hardware and software. Regarding software, it advises to "Employ anti-tamper technologies, tools, and techniques throughout the system
development life cycle" in its SR-9 Supply Chain Risk Management family of controls. Obfuscation is mentioned only as an option to strengthen the tamper protection, not to protect the original software. Moreover, the document does not discuss how to use these protections, or how to make decision regarding the selection of alternative methods. As such, NIST SP800-53 is not applicable to {\softprot}. In fact, for much of the remainder of this paper, we will actually discuss what a {\softprot} counterpart of NIST SP800-53 will need to entail. 

%\rthreenote{The paper should recognize and factor in frameworks such as NIST 800-53, BSIMM, OpenSaMM, OWASP ASVS.}\bdsnote{The latter 3 all relate to software security, more concretely to the development of software without flaws or bugs. The former also mentions reverse engineering, in relation to  TAMPER RESISTANCE AND DETECTION, but only refers to it as a functional requirement. No evaluation of how good such a technique is is discussed. SA-4,SI-7,SR-5,SR-9,SR11 in the document do discuss software integrity and software tampering, confidentiality of software assets is not mentioned at all, it concerns the protection of company assets and privacy against third parties or insider threats, not the MATE model. Obfuscation is also only mentioned with respect to tamper resistance of software.}
%\abnote{I only know OWASP ASVS, gues it is not addressing a MATE scenario. Attackers do not have full control of the execution environment but may have past knowledge of the source/binary code of the remote application they are tampering with. But if you can tamper with your copy locally you may also in some cases tamper with the remote version.
%NIST-800-53x is about the mitigations in the context of the 39. It can be useful to mention it should have a counterpart in the selection of the software protections to apply.}

%\abnote{add something about 800-53x}

\subsection{The state of \mate software protection}
Compared to network security and software security, the \softprot field has years of delay. For the sake of clarity, we reiterate from the introduction that the scope of {\softprot} includes protections that aim to safeguard the confidentiality and integrity of software by making reverse engineering and tampering harder. Table~\ref{tab:protection_examples} lists a number of well-known software protections. Out-of-scope of our work are mitigation techniques to prevent the exploitation of vulnerabilities in the software, such as \aslr, compartmentalization techniques, or safe programming language features in, e.g., Rust. Furthermore, it is important to understand that in the \mate attack model, attackers have full control over the devices on which they attack the software. They therefore can disable certain security features of the operating system and the run-time environment, such as \aslr. In other words, the execution environment cannot be trusted in the \mate attack model. 
For that reason, {\softprot} centers around protections embedded in the software itself, rather than relying on security provided by the run-time environment.

\begin{table*}
    \centering
    \begin{tabular}{lp{13cm}}
        \toprule
        protection type & explanation\\ 
        \midrule
            anti-debugging             & Checks whether a process is debugged (and stop executing if so) or techniques to prevent the attachment of an attacker's debugger, such as self-debugging~\cite{Abrath2020}. \\
            branch functions           & Replacement of direct control flow transfers by indirect, computed jumps to prevent reconstructing of control flow graphs~\cite{linn2003branchFunctions}\\
            call stack checks          & Control flow integrity checks that check whether called functions are from allowed callers, to prevent that functions are invoke out of their intended context \\
            code mobility              & Code is lifted from the binary to prevent static analysis on it. At run-time, the code is downloaded into the running application from a server~\cite{codeMobility}. \\
            code virtualization        & Code in the native instruction set (e.g., ARMv7 or x86\_64) is replaced by diversified bytecode and an interpreter is injected into the program to interpret the bytecode. As the bytecode format is diversified, it cannot easily be interpreted~\cite{Anckaert2006}.\\
            control flow flattening    & Control flow graph transformation in which a structured graph is replaced by a dispatcher that transfers control to any of the original nodes based on set data. This makes it harder to comprehend the original flow of control, and hence the functionality of the code~\cite{wangFlatteningTechReport}.\\
            data obfuscation           & Transformations that alter the values and data structures occurring in a the program to hide the original ones.\\
            opaque predicates          & Computations that evaluate to true or false based on invariants known at protection time (i.e., when the code is transformed) but hard to discover starting from a distributed binary~\cite{collbergOpake}. This enables the insertion of bogus control flow graphs, thus making comprehension and precise analysis of the code harder. \\
            remote attestation         & Techniques in which a remote server sends attestation requests to a running program, i.e., proofs of integrity. If the program fails to deliver a valid attestation, the program is considered to be tampered with, and an appropriate reaction can be triggered~\cite{viticchie2016reactive}. \\
            white-box cryptography     & Implementations of cryptographic primitives (e.g., encryption and decryption) such that even with white-box access to all data values occurring in the program state at run time, the used keys cannot be retrieved~\cite{wyseur2011white}\\
        \bottomrule
    \end{tabular}
    \caption{A number of software protections.}
    \label{tab:protection_examples}
\end{table*}

%Another reason to leave security features such as ASLR out of scope, is that their deployment is already standardized, and hence no longer the topic of a custom, application-specific risk analysis process.

The market of \changed{such} {\softprot} is neither open nor accessible to companies with a small budget.
In 2017 Gartner projected that 30\% of enterprises would have used \changed{\softprot} to protect at least one of their mobile, IoT, and JavaScript critical applications in 2020~\cite{Gartner}.
However, two years later Arxan reported that 97\% (and 100\% of financial institution) of the top 100 mobile apps are easy to decompile as they lack binary code protection or implement weak protection~\cite{arxan-report}. %, and the 100\% of the Financial Institution apps were easy to decompile, regardless of the presence of protections. 
A study confirms the absence of both anti-debugging and anti-tampering protections for 59\% of about 38k Google Play Store apps. The study highlights that weak Java-based methods are employed in 99\% of the cases where {\softprot} is applied~\cite{ceccato-new-one}. Malicious apps are obtained by repackaging benign apps to lure in victims \cite{Zhou2012repacked,Khanmohammadi2019repacked}. Repackaging is easy  because of the intrinsic weakness of the apk packaging processs but also because, when used, anti-repackaging protections are currently weak \cite{merlo2021repackage}.

%\bdsnote[inline]{TODO Aldo: fix the previous reference, that report does not contain information on the preceding statement} \abnote{I should have cited the wrong report, I didn't finf the one originally intended but I found this one https://arxiv.org/pdf/1902.00647.pdf, I'll look further}

Furthermore, the BSA Global Software Piracy Study~\cite{BSA} estimated that 37\% of installed software is not licensed, for a total amount of losses estimated as \$46.3 billion in the 2015-2017 period.
As a consequence, according to a Frost and Sullivan study~\cite{frost}, the \softprot Market, which accounted for \$~365.4M dollars in 2018, is expected to grow fast.
%, also easing the spread of malware that uses unlicensed software a major vector.

Cybersecurity competences are lacking~\cite{Gartner-report-online} and {\softprot} is no exception.
An Irdeto survey confirms that few companies have internal {\softprot} teams. Only 7\% of respondents stated their organization has everything it needs to tackle cybersecurity challenges, and 46\% stated they need additional expertise/skills within the organization to address all aspects of cybersecurity~\cite{Irdeto-report1}. Meanwhile many organizations lack competent staff, budget, or resources to protect their applications and systems~\cite{Mandiant}.

When the value of assets justifies it, developers resort to paying third parties to protect their software. 
However, the price tag is typically high, involving licenses to tools and often training by and access to expert consultants. Moreover, the services and the strength of the obtained \softprot are covered by a cloak of opaqueness. Security-through-obscurity is still omnipresent. For example, whereas early white-box cryptography schemes were peer reviewed~\cite{chow2002white,AESwhite} and then broken~\cite{DESbroken,AESbroken}, we could not find peer reviewed analyses of white-box crypto schemes currently marketed by big {\softprot} vendors. 
%Despite collaborating with industry, we as academic researchers get no access to their technical documentation, internal security analyses, and pentest report. 
%As another anecdotal piece of evidence, consider the fact that we, academic researchers that collaborate with commercial {\softprot} vendors in numerous bilateral and government-funded projects, hardly obtain any manuals of their commercial offerings or deeply technical documentation, security analyses, and penetration test reports thereof. In addition, 
Moreover, most vendors' licenses forbid the publication of reverse engineering and pen testing reports on their products, and they don't share their internal procedures, tools, and reports with the academic community. 
%As some of the big {\softprot} vendors are also active in other fields of (ICT) security, risk analysis and mitigation is almost certainly the principle that drives their {\softprot} experts. Yet a methodology for applying a risk analysis process for {\softprot} is not publicly available.
%
%Looking at the public records of attacks against some categories of applications that manage high-value assets and have devoted budget to {\softprot}, like media streaming services, online gaming, connected transports, and IoT, which resort to {\softprot} companies, we can observe that {\softprot} vendors seem to be effective in delaying attacks and potentially even in preventing attacks altogether, if only because attackers prefer attacking other software that is not well protected.

From this evidence, we deduce that either many organizations/companies do not understand the risk and therefore do not feel the need for deploying {\softprot}; or they do not have the internal competences and knowledge to deploy it properly; or they do not have enough money to pay third-party {\softprot} providers. In short, there exists no widely accessible, functional, transparent, open {\softprot} market. 
% driven by agreed upon standards and transparency.
%\abnote{the agreed upon standards is a bit forced}
At some of the big {\softprot} vendors, which are also active in other fields of (ICT) security, risk analysis and mitigation is most certainly the principle that drives their {\softprot} experts and that is encoded in internal policies. Yet a methodology for applying a risk analysis process when deciding how to protect is not publicly available.
Needless to say then, there is not a standard process that may guarantee the proper selection and application of available {\softprot}s given a case at hand.

\subsection{Promised land of software protection standards}
A methodological approach to risk analysis of software with respect to \mate attacks could have a plethora of benefits. In this section, we speculate on this potential.

%\rtwonote{On this whole section: None of these predictions are currently backed by fact or even logical arguments, but reads more like a marketing/VC brochure promising significant gains if this approach was supported. We would need more factual arguments motivating the particular approach}\bdsnote{Aldo, are there statements in here that you think we can back up with facts, e.g., similarities to what has happened in other fields. Alternatively, I propose to start the section by "could" instead of "would" and by an explicit statement that in this section, we speculate on potential benefits. That doesn't make the section stronger, but at least it makes it clear that we admit that this is speculative, hence taking away a stick from the reviewers.} 

First, it could force stakeholders to follow a more rigorous approach to \softprot. Indeed, risk framing forces the analysts to define workflows, processes, methods and formulas to evaluate risks and impact of mitigations. When applied to network security, a structured risk analysis has limited the impact of subjective judgments suggesting the use to collegial decisions.
% \abnote{something from ISO but anecdotal}
This could increase the transparency of all the protection phases, hence guaranteeing a more reliable estimation of the reached protection level and of the quality delivered by third parties.
In turn, we expect less reliance on the security-through-obscurity. 
Simply adopting the OWASP Security Design Principles forces software security specialists to avoid security-through-obscurity, which is also considered a weakness in MITRE CWE 656~\cite{CWE-656}.

Having a standard could finally induce the community to use well-defined terminology and to agree upon the meaning of each {term}, as happened after the NIST SP 800 documents for risk management~\cite{nistSP800-39}.

Building a common ground and well-defined playing rules would also benefit the {\softprot} market by creating a more open and transparent ecosystem where \softprot services can be compared as normal \emph{products}, {thus bridging the gap with the existing network security market, where products and their features are evaluated by third parties using standardized methods (e.g., Gartner Magic Quadrant for Network Firewalls~\cite{magicquadrant}).
Hence, we expect the raise of consultancy firms that can evaluate the effectiveness of {\softprot}s independently.
We would also expect a reduction of the prices of {\softprot}s and services in such a market, as highlighted by the Sullivan \& Frost study. %\abnote{no idea, trend of consultancy prices?} 
With a lower entry price and the definition of entry-level protection services, more companies will then afford professional \softprot services, with benefits for all the stakeholders including the final users.
}

%%%%%%%%%%%%%%%%%%%%%%%%%%%%%%%%%%%%%%%
When \softprot becomes standardized and the different aspects thereof hence become more clearly defined, it could also create a market for decision support products that automate the risk management. This could in turn lead to cost savings and more accessible and more effectively deployed \softprot. 

%\item 
As experienced elsewhere, the availability of standards increases awareness, as reported by an EU agency one year after the adoption of the GDPR \cite{gdpr1yafter}.
The simple existence of a standard would initially inform people about the need for protecting software.
Compliance would then force all involved parties to obtain in-depth knowledge about all relevant aspects. The standards and related best practices would therefore also be incorporated in educational programs. 
%Moreover, a standard would mitigate the risk for weak \softprot that is now evident in developers and companies that cannot afford services from vendors. \abnote{not sure about the message given but the last sentence}

The work towards standards could also have an impact on research. 
It would initially stimulate the community to focus on identifying and addressing the missing parts, and later on creating new or more effective and scientifically validated {\softprot}s that can be integrated into a standard framework and hence also becomes more reproducible. 
The interest in the field and the impact of work and publications would then likely attract more researchers in the {\softprot} field, which is now marginal in the software engineering community. 
We have found analogies with the impact of the ISO/SAE~21434 standard for cybersecurity engineering of road vehicles. 
Years before its adoption, car manufacturers have anticipated effort and funded research 
to cope with the new demanding standard. According to a McKinsey report~\cite{mckinsey-automotive}, the investments in automotive cybersecurity will grow from \$ 4.9 billion in 2020 to \$ 9.7
billion in 2030, with a market size expected to grow from \$ 238 billion in 2020
to \$ 469 billion in 2030.
%:2021 \abnote{something from the automotive of ICS field}
%https://www.mckinsey.com/~/media/mckinsey/industries/automotive%20and%20assembly/our%20insights/cybersecurity%20in%20automotive%20mastering%20the%20challenge/cybersecurity-in-automotive-mastering-the-challenge.pdf

%Moreover, after the task of \softprot is properly framed, the impact  on current SDLC practices needs to be evaluated, thus opening new research directions in software engineering. 
%\abnote{check swprot papers/conferences vs. software papers/conferences to state that there are just a few experts in sw protection}

% \pagebreak

An increased attention by research institutions and aca\-demia usually translates in better education opportunities, possibly with dedicated curricula on \softprot, which usually pair well with the career opportunities created by a more open market.
Ultimately, this could help companies to employ skilled people and support a freer job market that could partially compensate the lack for \softprot experts.

%\abnote{check trend in careers in AI and cybersecurity/network vs. number of experts needed.}

In the end, the benefit would extend to the whole community, as having better protected software reduces the global exposure of citizens to risks and, we hope, would make \mate attacks a less lucrative field, or at least reduce its growth.

\subsection{Challenges towards standardization}
\label{sec:challenges}
%\lrnote{Risk framing RQ12: adaptation of existing risk management procedures.}\bdsnote{will be done by integrating bullets that Aldo will provide in V.1) into previous sections.}

Despite the many benefits a standardized risk management approach would bring, such an approach currently is a long way off. Adopting a \nist-style risk management is a true challenge because we currently lack even the most basic common \mate vocabulary, but also because the assessment of exposure to threats needs to be tailored on the application to protect, and because the deployment of \mate\ {\softprot}s is in some regards more complex than securing networks.

%
% ASSETS
%
Today, we even lack a definition of asset categories. %
While the operations of a \mitm attacker are well defined, {\softprot} relies on the \mate attacker model that has never been defined formally. The abilities of \mate attackers are unclear, not in the least because formally modelling the fuzzy concept of code comprehension by humans and the ability to tamper with a program are more sophisticated than representing \mitm operations.
%
% ASSESSMENT
%
% \abnote{TODO: De Bello Gallico: done}
Estimating the feasibility of attacks requires a white-box analysis of the assets and of the entire application.
The complexity of mounting static, dynamic, symbolic, and concolic attacks depends heavily on the structure and artefacts of the software under attack, such as the occurrence of certain patterns in code and data or observable invariants in static code or in execution traces.

Unsurprisingly then, {\softprot}s as deployed in practice also provide only fuzzy forms of protection, not well-defined ones.
%As an instance, the structure of the program must be considered when estimating the effort needed to modify the application behaviour. \bdsnote{I propose to remove the last sentence above, it is too detailed.} \abnote{just from ``As an instance'' or the whole paragraph?} \bdsnote{the last sentence, so starting at "as an instance".}
%
% MITIGATION
%
%In the field of {\softprot}, mitigation is a complex task.
There currently is no well-defined set of categories of security controls to mitigate the risks. 
{\softprot}s have only been categorised in a coarse manner (e.g., obfuscation vs.\ anti-tampering). 
In general, it is not clear where they can serve and what security level they offer. 
By contrast, it is clear what firewalls and VPNs do and how to use them to mitigate network security risks. There {are} also accepted measures and guidelines to estimate the effectiveness of the different categories of network security mitigations and in some cases also categorization of tools and vendors that help in estimating the effectiveness of these mitigations~\cite{ISO27004}.

No metrics are available, however, to quantify \softprot effectiveness. While potency, resilience, and stealth are commonly accepted (theoretical) criteria that should be evaluated during a risk analysis~\cite{collberg1997taxonomy}, no standardized metrics are available for computing them. Complexity metrics originating from the field of software engineering have been proposed~\cite{D4.06} and ad-hoc metrics are used in academic papers \cite{JENS,linn2003branchFunctions}, but none have been empirically validated in the context of  {\softprot}, and none of them are trusted in practice, i.e., are deemed a sufficient replacement for human expertise and pen testing. 

%Importantly, protecting software is also more complex than protecting networks because {\softprot}s are not separate entities \changed{that each have their own logic and configuration languages and} that are deployed in front of or in between services and devices. \changed{{\softprot}s are in some cases not even entities, but processes.} 

%\rtwonote{Depending on the the specific software component or network, the opposite could also be argued (e.g., thinking about data vs. control plane discussions in global cloud provider or MNO networks, the whole next paragraph could be directly argued by replacing "software" with "network"). It would be important to argue why such statements should be assumed within the context of this article, and to be a lot more specific why they apply to SPs in particular (and not e.g. complex networks, which the authors argue are much better understood than software).}\bdsnote{Aldo, I think this is for you to fix, beyond the text in blue that I re-added but was commented out in the submitted version.}
%
\changed{Protecting software and networks are different tasks. While both come with their own complexities, there are aspects that appear more complicated in the protection of software. 
In software, the boundaries between assets and protections are not as clear as in the network scenario.
For instance, protecting networks often results in selecting the most appropriate  security controls and deploying them between attackers and assets (e.g., firewalls or VPN gateways to protect internal servers).
On the other hand, SPs are often processes that transform assets (e.g., pieces of code) to make analysis and comprehension of their logic more complex~\cite{schrittwieser2016}.
For instance, most forms of obfuscation transform code fragments. Since \softprot needs to be applied in layers for stronger and mutual protection and to exploit synergies, obfuscation can transform code that results from previous transformations or that has been added by other \softprot{}s (e.g., the code guards injected for anti-tampering purposes). Some obfuscations even aim for eliminating recognizable boundaries between different components~\cite{JENS}. 
As a result, the code of multiple {\softprot}s and of the assets they protect becomes highly interwoven. After their application, we may hence talk of protected assets, certainly not of separated protection and asset entities.}

\changed{Furthermore, for the deployment of {\softprot}s, software internals must be known to the tools, such as type of instructions, structure and semantics of the code, and the presence of any artifacts that might benefit attackers. This information is essential to decide if a (layered) {\softprot} may be effective or not, and to tune protection parameters.
In network protection, it is often enough to have black-box knowledge of the assets to protect to reach a reasonable level of protection (e.g., configure filters from the knowledge of authorized traffic to a server).
In addition, it is generally accepted that applications need to be designed with the protection of the sensitive assets in mind to deploy {\softprot}s effectively.
If the software architecture is not designed well (e.g., internal or external APIs misdesigned or authorization is missing between clients and servers), injecting {\softprot}s will only provide superficial mitigation of attacks. 
Therefore, risk assessment methods must recognize software whose design prevents proper protection and report that risks cannot be reduced to the desired objective by solely injecting {\softprot}s.
This scenario again stresses that the risk analysis of software applications requires insights into the internals.
A similar scenario happens with network services containing too many weaknesses that will later translate into vulnerabilities. These services will be exploited by attackers no matter the effort to protect them with security controls. However, in this case, there are tools for early detection of weaknesses, and statistics from vulnerability assessment can help correctly determine the risk profile.}
%\bdsnote{we still need to add something about the fact that a good report might help people to design software better. One questions is whether or not good design can easily be measured? For software engineering (maintainability, debugability, ... it can. But do we know of guidelines for secure software development for protecting assets against \mate attacks? By the way, we can point to Yuan's talk at Dagstuhl I think, and point out that at least standards are missing again. Or do secure design patterns and such help us out?}
%\abnote{I already added a sentence saying that risk analysis reports in D help in better protecting and understanding the picture, I forgot to cite Yuan's talk though.}

%

%Indeed, the processes associated to protection may be affected by the internal characteristics resulting in poorly protected applications. Again, it is peculiar the case of obfuscation, which in some case cannot hide semantics of programs or effectively hide sensitive data  \textcolor{red}{find interesting cases of data protection or the cases from Barak impossibility paper}.\bdsnote{I propose to remove everything above starting from "Indeed". That is too detailed I think.}

Moreover, in most  cases there are no hard proofs that {\softprot}s are actually effective in delaying attackers. Instead of encouraging checks from external parties, {\softprot} vendors prevent it, e.g., with legal contracts the analysis of the code after they have protected it. That is, they still rely on \emph{security through obscurity}. As a consequence there is neither an objective {n}or {a} measurable assurance that the software is well protected, nor an objective evaluation of the work made by these companies.
In academic {research}, the situation is not much better. For example, the seminal obfuscation versus analysis survey from Schrittwieser et al.\ ~\cite{schrittwieser2016} never refers to a risk analysis framework. Their results, although widely acknowledged to be very useful, can therefore not be immediately used in a decision process.
%Moreover, likely-invariants monitoring, a technique that has been proposed in literature for anti-tampering \cite{Kil2009RemoteAT}, has been later discovered as ineffective~\cite{viticchio2020impossibility}.

By contrast, in network security not only are protections considered to be separate elements, but assets are also considered to be black-box when performing a risk analysis, thus facilitating the way individual data are aggregated (e.g., the impact of different attacks or the evaluation of the overall risk status). While the identification of bugs/vulnerabilities in network protection components requires white-box analysis (which is, not accidentally, indeed an instance of an attack against software), bug exploitation as treated in a networking scenario does not consider the specific attack paths, only the likelihood that bugs may be exploited and their consequences, like in the case of zero days vulnerabilities, that are considered as incident or hazardous events, in no different way than earthquakes or flooding.

%\bdsnote{Can't we add a conclusion here, something along the lines of: in short, we lack standardization in SP (as discussed in part B), and that does not only result from being late, it also results from facing a much tougher problem than, e.g., the domain of network security? After such a conclusion, we can then move on to the next part, where we can actually start with the second part of the first paragraph of part B that I proposed to move, see my earlier comment, and with something along the lines of "Despite the huge challenges we face for moving towards standardization for the aforementioned reasons, we conjecture that ..."}
%\abnote{I agree a conclusion is really needed here. Moving the first sentences of section C is fine with me, but we have to check that the reader knows what we want to tell in the section (which, as you noted, is hard to read).}

In conclusion, despite their obvious appeal, risk management standardization and a functioning open market as they exist in other areas of ICT security are missing in {\softprot} not only because the community is late in developing them, but also because managing the risks is really challenging. 

%\abnote{having a risk analysis methodology/standard may highlight that if your application has not been designed to be protected the residual risks will be higher and maybe unmanageable, the awareness of these standards may help people in correctly designing their app since the beginning, swprot experts may be needed for this into the SDLC. How to redesign apps to be protected is out of scope.}

%

%\textcolor{red}{[REF IRDETO policy]}
%

%\section{\changed{Software Protection Risk Management: What?}}
\section{Proper Risk Management Requirements}
\label{sec:requirements}
% \label{sec:requirements}
In this section, we discuss what the four phases of the \nist IT systems risk management standard would entail as applied to {\softprot}, i.e., what tasks need to be done in each of the four phases, and what concepts and aspects need to be covered in them. We highlight some concrete challenges and discuss where we think existing techniques can be reused.

\subsection{Risk Framing}
\label{sec:framing}
The objective of risk framing is to define the context in which a concrete risk analysis will be performed. All relevant concepts and aspects need to be determined. To allow for standardization, a common vocabulary first needs to be established that covers all possible scenarios and relevant aspects. It needs to be unambiguous and formalized such that automated support tools can be engineered. In this section, we list the concepts we consider critical to frame the risks that {\softprot}s aim to mitigate. %It is for all of the discussed concepts and their various aspects that the unambiguous, standardized, and complete vocabulary and description methodology are to be provided such that risk can be framed for concrete cases at hand. 
Provisioning the complete vocabulary and a methodology to describe the risk frame is of course out of reach here. That will instead need to be done in a larger document that results from a community effort. %Later in the paper, we will provide some seeds, however.

\subsubsection{Assets} The \emph{primary assets} are all static and dynamic aspects of software of which a \mate attacker might violate security requirements because they have value for the attacker or the vendor of the software (monetary value, public image, customer satisfaction, bragging rights, ...). Some examples of primary assets are secret keys or other confidential data embedded in applications, algorithms that constitute valuable intellectual property or trade secrets and hence need to remain confidential, multiplayer game logic that needs to remain intact to ensure that players cannot cheat (e.g., see through walls, use aim-bots, and show full world maps), and authentication checks that need to remain in place in a program. These assets are the primary targets of \mate attackers. The assets cover a range of abstraction levels and granularities, that correspond to a range of software artifacts (functions, variables, global data, constants, etc.). For example, an algorithm that one wants to reverse engineer and steal can be large and expressed in abstract terms, while a secret encryption key to steal is merely a string of bits. 
Primary assets are typically already present in the vanilla, i.e., unprotected software.

The \emph{secondary assets} are all static or dynamic software artifacts that attackers might target as part of their attack path towards the primary assets. Those are the parts of the software that attackers handle on their way to their primary targets. Secondary assets can be attacker pivots in the vanilla software, but they can also be artifacts of injected {\softprot}s that attackers need to overcome to continue with their attack. An example of a pivot secondary asset is some encryption buffer that by construction contains high-entropy data. An attacker might first try to identify such a buffer in an application by applying statistical methods on memory dumps. Once the buffers have been identified, the attacker might pivot to the program slices that produce the data written in the buffer, and once those slices have been identified, the attacker can obtain the secret keys used in those slices, i.e., their primary target. An example of an injected {\softprot} is an integrity check. A gamer that wants to alter the speed with which he can move around in the virtual game world, might first have to undo or bypass the integrity check. 

The distinction between primary assets and secondary assets should not be made strict, however.  For example, a cryptographic key that protects one movie might be a secondary asset if the attacker is simply trying to steal one movie. A similar key that serves as a master key for all movie encryptions is clearly a primary asset. Moreover, \softprot vendors consider the {\softprot}s supported with their tools as being primary assets of which they do not want to see the logic reverse-engineered easily. While those protections are deployed to protect the primary assets of their customers' software, they are the primary assets of the \softprot vendors. Should attackers learn how to attack or circumvent those {\softprot}s automatically, their value goes down the drain.

For the deployment of certain {\softprot}s, it needs to be possible to describe the relationship between assets and non-asset program artifacts. This is the case when {\softprot} transformations applied to code of assets require other non-asset code to be transformed with it in order to maintain the overall program semantics. It is also necessary when deploying a \softprot on only the assets would make those assets' artifacts in the binary stand out to the attacker, e.g., because the entropy of encrypted data or obfuscated code is much higher than that of plain data or because the protection introduces recognizable fingerprints. To increase the attacker's effort needed to localize them, one can deploy the same {\softprot}s on non-asset code. In short, we need mechanisms to describe a wide range of software artifacts of assets, of non-assets, and of the relations between them. 

As {\softprot} aims to delay attacks rather than to prevent them, we need a way to describe the evolution of assets' values over time, as well as the impact that a successful attack can have on a business model. This includes the renewability of the assets, i.e., how easy it is to replace software to reduce the impact of successfully attacked assets. For modeling this evolving relation between business value and concrete assets, we expect companies to use their existing asset valuation models.

In the risk framing phase, the task for a case at hand is to determine which assets are potentially relevant to the risk management. This is needed for all the potential primary and secondary assets that are known a priori, i.e., in the original application, in already deployed {\softprot}s, if any, or in any of the {\softprot}s that might later be deployed in the mitigation phase.

\subsubsection{Security Requirements} The \emph{primary security requirements} for primary assets are most often the non-functional requirements of confidentiality and integrity, both of which come in multiple forms and at different levels of abstraction and granularity. The requirements' scope differs from that in other domains, and existing classifications can hence not be trivially reused. Most importantly, \mate integrity requirements can include constraints on where or how code is executed. Examples of such requirements are that at any point in time, at most one copy of a program is running, or that certain code fragments are not lifted from a program and executed ex situ. 

Different requirements may hold for different phases in the \sdlc of the software to be protected and of attacks (e.g., attack identification phase versus attack exploitation phase). Some requirements may be absolute (e.g., some master key should not be leaked at all), others may be time-limited (e.g., a key to a live sports event should remain secret for 5 minutes), still others may be relative and economic (e.g., running many copies in parallel undetected should cost more than licensing them).

Evaluating whether or not such primary requirements can be guaranteed is extremely hard in practice because \mate attackers have white-box access to the software and full control over their devices. \emph{Secondary security requirements} can then help for framing possible risks. The secondary security requirements can be (i) non-functional requirements for secondary assets; (ii) functional requirements that are easier to check but of which the mere presence in itself provides few guarantees, such as the presence of a copy-protection mechanism; (iii) assurance security requirements that minimize the risk that relevant aspects are overlooked; and (iv) what we will call \emph{protection policy requirements}. The latter are requirements related to worst-case assumptions about attacker capabilities, such as assuming that the mere presence of some artifact or feature suffices to enable certain attacks. Such assumptions to some extent allow to make up for the lack of proper evaluation of primary requirements. For example, a lack of stealth resulting from easily identifiable invariants in injected {\softprot}s hints for potential weaknesses vis-\`a-vis certain attacks~\cite{yadegari}. Protection policy requirements then come down to requiring that certain forms of artifacts and features are not present at all. This is similar to security policies in the domain of remote exploitation, where, e.g., code pointer integrity~\cite{CPI} is a policy about handling code pointers that, when implemented properly, ensures that control flow cannot be hijacked by exploits.

In the risk framing phase, the task for a case at hand is to determine the security requirements for all assets and potential weaknesses identified as relevant as discussed above.

\subsubsection{Attacker Models} It needs to be possible to consider and define attackers with different levels of resources and capabilities; money, expertise, available tools, etc. The latter involve a huge range of methods, span multiple levels of abstraction, and evolve over time, so a \emph{living catalog} is needed. We currently do not know what level of detail will produce the best results, so both more generic attack methods and tool usage scenarios (e.g., disassembling code) and very concrete ones (e.g., using the IDA Pro 6.2 disassembler) need to be covered. As the goal of \softprot is to delay attacks, not only the feasibility of successful attacks is to be covered, but also the potential effort involved, possibly including the effort and time that attackers would probabilistically waste in unsuccessful attack strategies. 

While research has shown that unsuccessful attack steps are common in real attacks~\cite{emse2019}, it is unclear whether attack models that assume worst-case scenarios can be useful or even sufficient. Worst-case assumption examples are attackers being served by an oracle to always choose the right attack path, and analysis tools producing results with ground-truth precision. For example, locating code of interest is an important, time-consuming attack step that cannot simply be assumed to be performed effortlessly by means of an oracle. Doing so would imply that increasing the stealth of {\softprot}s is not useful, which experts certainly reject. Importantly, for each potential attack step considered for a concrete case at hand, it needs to be possible to clarify in clear terms what enables or prevents the attack step (e.g., the presence of certain secondary assets), and what significantly increases or decreases the required time and effort, and the likelihood of success. This might include features of the software under attack, of the environment in which attacks can be performed, but also knowledge obtained by the attacker. The best abstraction levels to consider these is an open question. 

In the risk framing phase, the task for a case at hand is to determine the attacker model precisely, i.e., the different combinations of the mentioned attributes that potential attackers in scope might have. Once the attack scope has thus been determined, it is important to frame the set of (quantitative) metrics that will be considered useful in the later phases to approximate the effort/time/resources that attackers will need to invest in the different attack steps in scope. As already discussed in Section~\ref{sec:challenges}, there is no widely accepted set of metrics at this point in time. More scientific, empirical research is needed to determine which metrics are valid under which circumstances and for which purposes.

Existing attacker models from, e.g., network security risk analysis cannot be reused. \mate attack modeling needs to include manual tasks and human comprehension of code, which are not considered in network security. 
For example, in network security, the development of zero-day exploits (using tools also found in the \mate toolbox) is handled as an unpredictable event, which side-steps the complexity of analysing and predicting human activities. This entirely prevents the use of existing assessment models developed for the network security scenario.

\subsubsection{Software Protections}
It needs to be possible to describe a wide range of available {\softprot}s in a unified manner. This needs to include at least possible limitations on their applicability and composibility, be it for layered deployment to protect each other or in another way to exploit synergies between multiple {\softprot}s; the security requirements that they can help to enforce; (measurable) features or limitations they have that can enable, slow-down, ease, block, or otherwise impact potential attacks at different levels of abstraction, on the {\softprot}s themselves but also on the assets they are supposed to protect; how big those impacts are on the different potential attacks; and potential implementation weaknesses including the ways in which they can fail to meet secondary security requirements, i.e., become (easily) attackable assets themselves; etc. The link to validated (but as of yet still missing) metrics mentioned above is clear. 

Also the costs of using a \softprot need to be considered. This can include the direct monetary costs of {\softprot} tools licenses, but also indirect costs such as having to budget for more security servers or having a longer time to market, or any other extra cost that might follow from required changes to the \sdlc. 

In addition, the amount of potential overhead in terms of application performance (latency, throughput, size, ...) needs to be known of all potentially used {\softprot}s. This is critical, because many applications have little overhead budget when it comes to responsiveness, computation times, etc. Part of the performance impact depends solely on the {\softprot} itself, such as the time or memory required to initialize a {\softprot} component. Other parts of the impact can depend heavily on the specific way in which the {\softprot} are deployed. For example, whenever a protection requires the injection of a few instructions into original code fragments, the resulting overhead will depend heavily on how hot (i.e., how frequently executed) that fragment is. Multiple ways for expressing the potential cost of {\softprot}s are hence needed.

In the risk framing phase, the task for a case at hand is to determine precisely which combinations of {\softprot}s can \emph{potentially} be deployed to mitigate risks, given the available {\softprot} tool flow and the relevant properties of the supported {\softprot}s.

\subsubsection{Software Development Life Cycle Requirements}
\label{sec:framing_sdlc}
Besides mitigating attacks, {\softprot}s come with side-effects, such as slowing down software, making it bigger, making debugging harder, requiring changes to distribution models, requiring a certain scalability on the side of secure servers, etc. Taking the time to decide on {\softprot}s, possibly iteratively with the involvement of experts and time-consuming human analysis, also has an effect on the time to market.

Before choosing the most effective mitigation in the risk mitigation phase, hard and soft constraints need to be listed with respect to quantifiable overheads (i.e., costs) in all possible relevant forms, and with respect to compatibility with \sdlc requirements. Note that different constraints might apply to different parts of a program. For example, in an online game or when streaming a movie, the launching of the game or player might have a large overhead budget, while during the game or movie, good real-time behavior is critical. 

For all available {\softprot}s, later phases of the risk analysis will need to be able to estimate the impact on the relevant costs and \sdlc. It is therefore necessary to obtain all relevant profile information about the software to be protected, including execution frequencies of all relevant code fragments. 

An issue that complicates matters immensely in practice, is the fact that often vendors of \softprot tools (hereafter named \softprot vendors) and users of such tools (hereafter called application vendors) do not trust each other. Both parties hence often put severe constraints on the way the \softprot tools are deployed on the applications and on the amount of information that they exchange. A \softprot vendor will typically not be very forthcoming with respect to the weaknesses or internal artifacts of the supported {\softprot}s and disallow reverse engineering of them, while the application vendors do not want to share too many internals or code of their software with the \softprot vendor. Consequentially, only illegitimate attackers will get white-box access to the protected applications in which {\softprot}s and original assets are interwoven as discussed in Section~\ref{sec:challenges}. If the experts performing the risk management lack white-box access to all available {\softprot}s and to the protected application, this will have a tremendous impact on the methods and data that can be used during the risk assessment and risk mitigation phases in which attackers with white-box accesses are targeted. This obviously needs to be documented, and the impact thereof needs to be assessed, during the risk framing.

%As for the available experts, their involvement is so important that we think it is necessary to explicitly consider their availability during the risk framing phase. \softprot is so complex, and available expertise so scarce and expensive, that applying the best practices for all of the different risk management phases might be beyond the reach of an application vendor. During the risk framing phase, it is therefore important to consider which expertise is and will be available and which will not be available, such that the later phases can be adapted, e.g., by excluding the use of certain \softprot schemes/tools or by opting to use simpler, more worst-case scenario assumptions than results of complex analyses, and such that the impact of the lack of expertise can be considered.

In addition, aspects of the \sdlc relevant to the monitoring phase that will be discussed later need to be framed, such as connectivity and updatability. Whether an application is (or can be required to be) always online, occasionally connected, or mostly offline impacts which online protections can be deployed, and hence which monitoring techniques will be available. So does the ability to let application servers (e.g., video streaming servers or online game servers) interact with online security services such as a remote attestation server. Likewise, it is important to document whether updates can be forced upon users, and to what extent the vendors can synchronize all user updates.

Finally, limitations to the environment in which protected applications will be distributed and executed need to be documented. For example, Android allows less freedom than Linux regarding the use of certain OS interfaces, e.g., for debugging, and some device vendors limit what applications can do after being installed, such iOS's limitation on downloading binary code blobs post installment. Such limitations clearly affect the types of {\softprot}s that can be deployed, so they need to be included in the risk framing.

To avoid the need for costly human expertise and manual intervention in the following process step, as much as possible information discussed above needs to be formalized, such that tools can reason about them in the subsequent phases.

%\lrnote{Risk framing RQ9/13/14: usability by {\softprot} experts/non-SWprot-experts-application-developers} \bdsnote{I think those are useful questions to ask in general, but in this paper, we already assume positive answers, don't we?}

%\todo[inline]{collect what can be reused}

%%%%%%%%%%%%%%%%%%%%%%%%%%%%%%%%%%%%%%%%%%%%
%%%%%%%%%%%%%%%%%%%%%%%%%%%%%%%%%%%%%%%%%%%%
\subsection{Risk Assessment}
% \abnote{discuss impact of servers used by \softprots in the overall risk}
\label{sec:assessment}

In the above discussion of the risk framing phase, the term ``potential" occurred frequently. The reason is that in the risk framing phase, all forms of knowledge are still considered in isolation, including potential \softprot weaknesses, application features, \softprot tool capabilities, and attacker capabilities. Assessing how those interact for the case at hand, i.e., determining which of all potential risks actually manifest themselves in the software at hand, is the goal of the risk assessment phase.

So first concrete threats and risks need to be identified, starting from an analysis of the assets, their intrinsic weaknesses, and from attack methodologies, how they are instantiated, i.e., their technical attributes that impact their feasibility.
Then, a qualitative, semi-qualitative, or preferably a quantitative estimation of the impacts of these threats and a prioritization of the risks needs to be done.

%
%Threats against the assets need to be considered, in the framed attack scenario and the way weaknesses could be exploited to affect security requirements. 
%
% \bdsnote{In the above, what is the difference between the "first" part, and the "then" part? They look identical to me...}

%In some cases, protections may also be part of the design of the application. 
%For instance, when cryptographic keys are primary assets, application designers may have already foreseen the use of white-box cryptography solutions. 
%Therefore, the risk assessment phase must be able to reason about applications and assets that may already incorporate protections.
%Moreover, in the \mate domain, everyone agrees that protections should be combined and layered on top of each other. \abnote{a bit too anecdotal} 
%In order to decide on the need for additional layers or to determine that no more layers are needed, it therefore needs to be possible to assess the risks on already protected applications. As a consequence, all methods and tools for performing the risk assessment need to be applicable to applications protected to varying degrees, ranging from completely unprotected to heavily protected. 

%Therefore, evaluating the impact of the threats requires \textcolor{red}{ad hoc techniques/more careful operations.}

%\abnote{check flow and add glue}

%\abnote{we should elaborate a bit more on layered protection, as a significant difference in the approach, maybe move this paragraph elsewhere}

%\abnote{after one layer the protections are interleaved with original code}

%%%%%%%%%%%%%%%%%%%%%%%%%%%%%
\subsubsection{Identification of the threats} %Assets \& Security Requirements}
\label{sec:identification_threats}
% \abnote{attack strategies (static vs.\ dynamic vs.\ hybrid vs.\ traditional vulnerabilities and bugs}\\
% \abnote{attack tools}\\
% \abnote{locate first attack then}

The goal of this phase is to determine a list of the attacks that, if executed successfully by an attacker, would lead to the violation of security requirements expected on each software assets.
Therefore, this phase consists of a detailed analysis that outputs a report about the risks in terms of the analyzed attacks that are viable within the relevant time frame (in which assets have value and security requirements), the attack paths of least resistance, the levels and amounts of expertise, effort, and resources they need, the damage caused by exploitation, etc. 
For each attack path contributing to the major risks, the weaknesses and secondary assets used by attackers as pivots need to be presented, as well as the used assumptions, such as worst-case-scenario considerations or parameters that are unknown in practice. Reporting this information is necessary to enable confidence in the outcome of the assessment. 
Critically, all of the enumeration and assessment of feasible attack steps must be performed both on the attack identification phase (in the attacker's lab on devices controlled by attackers) and on the attack exploitation phase (often outside their lab, on other user's devices). 

Several open issues need to be addressed to perform this identification task correctly in the \mate scenario.

Software assets can be attacked with different strategies, in which attackers rely on automated tools and analyses to collect and exploit information about the software under attack and to represent the software in structured representations. A range of analysis tools and techniques are applicable, all with their own strengths and limitations, including static, dynamic, symbolic, and concolic analyses. % The analyses produce information about the software that the attacker is seeking to exploit, possibly to enable further attack steps. %The results All results from the analyses provide  about the software, the goal of the attacker using the tools is to collect those facts. 
% The identification of the attack techniques and strategies strongly relies on the knowledge of the tools that attackers have at their disposal.
Knowledge of the attacker's goals and tools is the starting point to identify and enumerate the possible attack paths. This knowledge includes the kinds of information the different tools can produce, the software features they depend on to produce that information, their weaknesses, limitations, and precision. 

In this phase, the defender therefore needs to deploy his own analysis toolbox to determine the technical attributes of the primary assets that are present and that have an impact on the risk of successful attacks because they enable them. This at least includes checking whether the protection policy requirements formulated in the risk framing phase are violated. It also needs to be done for all potential weaknesses that were identified in the framing phase, such as invariants being present in the code that might facilitate certain attack vectors, or other features that result in a lack of stealth or that can serve as proxies for concrete attack opportunities. Moreover, the set of actually present secondary assets needs to be determined to identify the presence of features that make them good pivots for attackers towards the primary assets. 

While we are convinced that such defender toolboxes can produce most of the necessary information about the feasible attacks to be enumerated, a number of research questions still need answers. For example, how can the formal pieces of information extracted by the tools be used to identify the viable attack paths precisely, in particular when attackers need to resort to manual efforts that may not be easy to formalize. How do we then assess the required effort and likelihood of success, and perform impact analysis and risk estimation accurately? And to what extent can such automated analysis with a defender toolbox suffice to avoid the need for actual penetration testing involving human experts? %It would be interesting to determine how this knowledge can be used to automate attack identification. 

%%%%%%%%%%%%%%%%%%%%%%%

How fine-grained or concrete the enumeration of considered attacks paths needs to be is also an open question, with respect to how paths are composed of attack steps and how their attributes are aggregated, and with respect to how concrete individual attack steps need to be.
Since the assessment must drive the mitigation, the generated information must be rich enough for the mitigation decision makers. 
Therefore, to some extent, the answer to this question  will depend on the goal of the assessment, which might be, e.g., a semi-automated or fully automated mitigation phase. In the latter case, assessment information must be extensive and accurate, as an automated decision support system cannot rely on human intuition and experts' past experience.

%%%%%%%%%%%%%%%%%%%%%%%%%%%

The identification of attacks with an analysis toolbox requires white-box access to the application code; a black-box approach as in computer networks is not viable. In case white-box access is not possible, e.g, because of \sdlc requirements discussed in Section~\ref{sec:framing_sdlc}, alternative sources of information about the features and weaknesses present in the different integrated components need to be considered, such as partial analysis reports provided by the involved parties. Alternatively, and as long as the discussed enumeration approach cannot completely replace human expertise, the inclusion of results of penetration tests performed by so-called red teams could be considered.  %Possible weaknesses in the design and implementation of the application to protect seriously affect the attacks that can be mounted and need to be identified.
In short, the risk assessment phase needs to be able to take into consideration a wide range of information sources and forms.

For the scalability and practical use of a software risk analysis process, another open issue is how to update and maintain the attack path enumeration without repeating a full analysis from scratch when any of the involved aspects evolve while the application is still being developed, be it the application itself, the protection tool flow, the attackers' tool boxes, etc. Especially if the enumeration of the attacks involves human expertise, this maintenance issue is critical to keep the risk management approach to \softprot viable.   

With the current state of the art, such human expert involvement is still necessary. Past research aimed to automate the attack discovery with abductive logic and Prolog~\cite{reganoProlog}. That suffers from computational issues, since generating attack paths as sequences of attack steps causes a combinatorial explosion and requires massive pruning.
With the pruning implemented by Regano et al., only high-level attack strategies can be generated, which often do not contain enough information to make fine-tuned decisions when similar {\softprot}s are considered. For example, they allow determining the need for using obfuscation but do not provide hints for selecting among different types of obfuscation.

Several results in the machine learning field are potentially applicable to synthesize attack paths from attack steps in a more effective way. 
For example, methods behind exploit generation~\cite{angr,brumley_apeg} techniques that automatically construct concrete remote exploit payloads to attack vulnerable applications, could be investigated to check if aspects of it are suitable to determine \mate attack paths automatically. 
Certainly, they will need modifications, as finding remotely exploitable vulnerabilities is rather different from finding \mate attack paths that violate security properties of assets.
%\bdsnote{I find this a "dangerous" paragraph. Why suddenly start comparing to a completely different attack scenario (remote exploits)?}
%\abnote{rephrase to highlight differences and evaluate if after lifting is worth to stay in the paper. ML may have produced usable results for synthesis. prune not relevant, prune this space similar techniques to AEG may help}

% The objective of risk framing is to define the context in which a concrete risk analysis will be performed. All concepts and aspects relevant to the risk analysis need to be determined. To allow for standardization, a common vocabulary needs to be established a priori, which is complete in covering all possible scenarios and all their relevant aspects. It needs to be unambiguous, and ideally it is

For example, in the \mate risk analysis methodology, for each identified attack path, defenders need to estimate the likelihood of succeeding as a function of the invested effort, attacker expertise, time, money, luck in trying the right strategy first or not, etc. All of that is absent in the mentioned automated exploit generation.  
%
%Estimating the information that an attacker can gain using analysis tools is crucial in this phase, as well as models to determine how this information is usable to mount further attacks, including understanding when attackers have to resort to manual effort instead of automatic exploitation.

Regarding automation of this phase of the approach, we think the identification and description of primary assets cannot be automated, as those assets often depend too much on the business model around the software. They can hence not be derived from analyses of the software alone. By contrast, the identification of secondary assets, as well as the discovery of attacks paths and the assessment of their likelihood, complexity, and other factors that contribute to the overall risk, should be prime targets for automation. Even if full automation is out of reach because parts cannot be automated or some parts do not produce satisfactory results, automating large parts of the threat identification phase will already have benefits. It will reduce the need for human effort, thus making proper risk assessment cheaper and hence more accessible, and it can raise awareness about identified attack strategies, thus making the assessment more effective. A gradual evolution from a mostly manual process, over a semi-automated one, to potentially a fully automated one, is hence a valuable R\&D goal. We stress that in order to succeed, the automated tool support should then not only provide the necessary inputs for later (automated) phases of the risk management, it should also enable experts to validate the produced results to grow confident in the tools. Section~\ref{sec:workflow} will present a tool that, although being rather basic, achieved just that.

% After identifying the type of information, quality and usefulness of the tool information need to be assessed, in order to determine how easy is for automatic tools and human beings to use tool information to attack software. 
% The effort needed to comprehend the information output by tools needs to be considered. 
% This comprehension affects two phases, asset localization and tampering. \bdsnote{I am lost here....}

\subsubsection{Evaluating and prioritizing risks}
% \abnote{explain the consequences and side effects of assets violated }

The risk assessment report must give an indication of the consequences that exploitation of the risk may have. It must produce an easily intelligible value or score associated to all the risks to all assets. 
Since the objective of the report is prioritizing the risks in order to drive the mitigation phase, it must not only consider the value of the violated primary assets, but also the side effects, like impact on business reputation or market share losses.

Furthermore, it may consider the likelihood that attackers are interested in executing the identified threats because of different expected {\roi}s. For example, an attack path that offers a lot of potential gains for the attacker (or damage to the company) might be less attractive when it comes with a high probability of being detected and having to face legal consequences.

{
When outcomes from the impact analysis are available in proper form, our feeling is that this phase has no peculiarities compared to risk analysis in other fields. 
Models can therefore be adopted from existing literature to build a system that allows the consistent evaluation of the impacts. 
As a promising option, we consider risk monetisation~\cite{doerry2015monetizing}, the process of estimating the economic loss related to a risk and the \roi of a mitigation activity, which eases the reporting of risk assessment data to the higher management and is general enough to work for every asset type, including software assets.
As another instance, aspects of the OWASP risk rating methodology for web applications might be useful~\cite{OWASPrisk}. Automation support for the available options can then obviously also be reused, possible after some adaptations.
}

%\lrnote{I think a paragraph to relate with existing risk assessment techniques (see Risk Assessment RQ8) could be useful, especially to link after with the usage of attack paths in ADSS}

% \abnote{collect what can be reused from net sec}

\subsection{Risk Mitigation}

This phase comprises two parts: (i) mitigation decision making, and (ii) implementing and validating the decisions. 

\subsubsection{Decision making}
\label{sec:decision_making}
First, one needs to evaluate how the deployment of certain combinations and configurations of {\softprot}s with will affect the high(est) risk attack paths determined before. %, either by preventing the execution of those attack paths as is, instead requiring additional or alternative attack steps; or by requiring attackers to invest significantly more effort, time, and resources to execute them. 
Ideally, this evaluation can be done through estimation, i.e., without having to actually deploy the considered {\softprot}s and having to measure the effect of that deployment. This is a major difference with the risk assessment phase, which relied heavily on measurements. How precise the estimations need to be to enable sufficiently precise computation and comparison of residual risks, is an open question. We hence consider two possible approaches for this process. 

A first approach builds on the assumption that estimations are sufficiently accurate to determine the best possible combination and configuration of {\softprot}s without requiring any measurement. One then first determines the combination and configuration that achieves the minimal residual risk while not violating any of the hard constraints. Next, one select alternative targets that trade off some of the residual risk for other aspects, such as lower costs in the form of performance, required adaptions of the \sdlc, user-friendliness, etc. For each alternative target, one then again selects the best {\softprot}s and estimates the delta in residual risk and in the other relevant aspects over the selection that yielded the minimal residual risk. Finally, one then makes a choice between the most protective selection and the alternatives. This human decision will typically involve {\softprot} experts, application architects, and company managers familiar with the overall business goals and strategy. 

Given the complexity of {\softprot} as discussed above, i.e., the fact that {\softprot}s are not separate black-box entities, we consider the above decision making process not viable at this point in time and in the near future. It is simply out of reach for humans and for current decision support tools.  

An alternative approach, familiar to practitioners in industry, is to add additional {\softprot}s iteratively in a layered fashion. The assessment and mitigation phases are not executed once, but alternating over multiple rounds. In each round, an assessment phase is followed by a mitigation phase. In the first round, the risk assessment phase is performed on the vanilla application. In later rounds, the assessment phase is performed on the version of the application protected with all {\softprot}s selected in previous rounds. During such later assessment phases, measurements are performed on already selected and deployed {\softprot}s. This provides a workaround for the lack of good enough estimation methods as needed for the first approach. 
%Additional risks introduced by the deployment of {\softprot}s need to be considered as well in estimating the residual risk. For instance, remote attestation techniques rely on the presence of a trusted remote server that checks software integrity. While research papers assume these server as trusted and invulnerable, considering risks in the real world is different and attackers controlling these servers may hide their activities. 
It also eases the handling of novel risks introduced by deployed {\softprot}s, such as when the location of non-stealthy {\softprot}s might leak the location of the assets.

In each round, the mitigation phase adds a limited number of additional {\softprot}s to the ones already selected in the previous rounds. In each round, different combinations and configuration of {\softprot}s can be proposed that offer different trade-offs in terms of risk reduction and costs. Humans will then again select one set of {\softprot} configurations and continue to the next round, or stop once the whole budget in terms of costs is consumed or no more significant risk reduction is achieved. In each round, different constraints are imposed that limit the {\softprot} configurations considered in that round, and the set of {\softprot} configurations is chosen that meets those constraints and that offers the best potential to reduce the residual risk. Estimating the reduction potential rather than the immediate reduction in each round (except for deciding whether a round will be the last round) allows for taking into account a priori knowledge about the fact that some {\softprot}s have the potential to become much stronger after additional rounds (i.e., additional layers) are deployed, while others cannot become stronger because of a lack of synergies. An example of constraints evolving between rounds is that in the first rounds, {\softprot}s might only be deployed on assets, while in later rounds non-stealthy {\softprot}s can become deployable on non-assets to avoid that protected assets stand out because of protection fingerprints as discussed in Section~\ref{sec:framing}. Our \poc presented in Section~\ref{sec:workflow} contains such an asset hiding round, as will be detailed in Section~\ref{sec:esp:workflow:hiding}.

The alternative approach is more realistic for several reasons. The humans making decisions in each round can make up for deficiencies in the existing tool support and formalized knowledge, and can build more confidence in the outcomes of the mitigation process. Secondly, measurements are performed in each round, which again allows for more confidence in the outcomes. 

It is on the mitigation task that automation poses the most severe constraints. 
Optimizing the selection of the combination of {\softprot}s to deploy must comply with computational constraints. In most usage scenarios, optimization models must guarantee to return results in minutes or hours. Given the large search space to explore, this implies the need for ad hoc models that prune the less relevant combinations efficiently. In some usage scenarios, it can be acceptable that the optimization models return far-from-optimal results quickly, such that the time-to-market requirements of an initial software launch can be met, while spending more time to find better combinations of {\softprot}s for updates of the software.\footnote{Anonymously, \softprot suppliers confirm to us that for many of their customers the norm is weak implementation at first because security/protection is not on the feature list required by product management, and then complaining when things get broken, after which the \softprot supplier needs to help out. Obviously those customers prohibit documenting concrete cases.}

Within one round of decision making, the optimization process should be driven by at least the potency of the selected combination of protections and by the estimation of the performance of the protected app (\eg, user experience). Ideally, resilience is also considered. Current methods for assessing the potency, resilience, and (to some extent) overheads are not usable for automatic decision support, however, as they require the actual application of the protections to perform a measurement on the protected version. Given the time and resources needed to apply protections on non-toy programs to measure objective metrics, and to measure the overheads by running the protected applications, an optimization process that requires measurements instead of estimations would only consider a very limited solution space, which would make the optimization process useless.

Estimating the strength (and overhead) of {\softprot}s is really hard, however, given that {\softprot}s are composed and layered, resulting in their code being highly interwoven. Methods from, e.g., network security to aggregate the strength and overhead of combined but clearly isolated network security controls are hence not reusable. Machine learning techniques can possibly solve this difficult problem, but clearly need further research. 

Another open issue is that different {\softprot}s have different effects on attack success probability, in particular when the security requirements are time-limited or relative. In some cases, the effects can be quantified in absolute terms, such as increased brute-force effort required to leak an encryption key from a well-studied white-box crypto protection. In other cases, such as the delay in human comprehension of code that has undergone design obfuscations~\cite{collbergbook}, the effect is harder to quantify precisely and in absolute terms. When software contains different assets with different forms of security requirements, the relative value of different protections hence becomes very difficult to determine, and hence the overall risk mitigation optimization becomes increasingly difficult.

%\changed{As a solution, we have applied ML techniques to estimate the changes on the objective metrics after the application of protections. From the predicted metrics we compute the new potency. Also our simplified overhead estimation model is based on predicted metrics and data about the protections, nonetheless a new model that uses ML is under development.} \bdsnote{I don't like the way this blue text is phrased yet. Has this been published? Is this part of the ESP later in the paper? Either way, the text has to be rephrased to make this more clear.}

\subsubsection{Actual deployment}
In each mitigation round, the chosen combination/configuration needs to be deployed, i.e., {\softprot} tool flows need to be configured and run on the application to inject the {\softprot}s selected in all rounds so far. 

Ideally, this part is completely automated. This obviously requires communication interfaces between decision support systems and protection tools. Providing such interfaces and enabling this automation would have significant benefits. Apart from saving effort on manual interventions for protecting concrete application, it would also skip the learning curve of how to properly configure the deployment of protections with specific tool flows. Moreover, having such an integrated framework could pave the road for an open standard for an API for software protection.

Following the deployment, it is critical task to validate that the {\softprot} tools actually delivered as expected. Were the selected {\softprot}s injected in the intended way? Do the injected {\softprot}s have weaknesses that were not expected? How to obtain the necessary validation is an open question in some usage scenarios, in particular when the deployment of the mitigation is executed by multiple parties that don't want to share sensitive information. % To automate this validation, available tools of the risk assessment can most likely be reused, such as the analysis toolboxes.

%%%%%%%%%%%%%%%%%%%%%%%%%%%%%%%%%%%%%%%%%%%%%%%%%%%%%%%%%%%%%%%%%%%%%%%

\subsection{Risk Monitoring}
% \abnote{Aldo will write bullets, now that we will include remote attestation in the figure of the workflow}

% TODO offline/online monitoring?
% TODO: which actions should be linked to offline monitoring?
% TODO: additional assets in original version versus new program version

According to the NIST~\cite{nistSP800-37} risk monitoring includes ``assessing control effectiveness, documenting changes to the system or its environment of operation, conducting risk assessments and impact analysis, and reporting the security and privacy posture of the system.''
In the \softprot scenario, monitoring involves the continuous tasks that need to be performed once the protected software has been released, in order to keep track of actual level of exposition to the risks over time. 
This consists of two related activities: keeping the risk analysis up-to-date, and keeping track of the risk exposure of the released application. %The former is related to howthe evolution over time of the previous three activities, whereas the latter is related to monitoring the concrete risks that occur in the instances of the protected application running on the premises of users.

\subsubsection{Keeping the risk analysis up-to-date}
% the differences are how monitoring data are collected: deploy
% remote vs. offline: change the control in the framing

The task of keeping the risk analysis up-to-date tracks how the inputs used in the past risk analysis activities of framing, assessment, and mitigation evolve over time, and how that evolution affects the decisions that were made in those activities. We can abstract these into monitoring the evolution of three different pillars of information: the information related to the assessments (\eg, new attacks, attack techniques, tool updates), the information related to {\softprot}s (\eg, updates, vulnerabilities, breaches), and the information related to the protected application. 
Of course, monitoring can then lead to the decision that a differently-protected version of the protected application should be released whenever any tracked changes lead to the need to re-evaluate earlier decisions as to which risk are prioritized and which mitigations are deployed.

The monitoring of the information related to {\softprot}s involves keeping track both of attacks against existing {\softprot}s and of newly developed {\softprot}s. For example, when a complex attack technique (e.g., generic deobfuscation~\cite{yadegari}) is first presented in the academic literature, it might not be considered practically relevant during an original risk assessment because the attack is hard to replicate and its effectiveness has not been demonstrated on more complex pieces of software. However, when attackers later release a tool box that automates the replication and they publish a blog discussing how they used it to attack a complex application successfully, this should lead to a re-evaluation. Similarly, when new {\softprot}s become available with higher effectiveness against old or new attacks, or with lower overhead, this may lead to a re-evaluation.

%Other changes in the {\softprot} landscape that can affect the risk assessment are new {\softprot}s that become available on the market, with different properties than those that were considered before. For example, they might protect against certain attacks in a much stronger fashion or with significantly less overhead than previous techniques. All these changes in knowledge of the {\softprot} landscape can then lead to the knowledge that changes need to be made to earlier decisions.

Similarly, the information related to the application that was used as input for the previous tasks can also evolve over time. One example is where a company might decide that there are, in fact, additional assets in the program that need to be protected. This can happen both as a late realisation after deployment, but also in the case where the application itself evolves over time, by virtue of new versions being released with changes in functionality or structure. Another example is that the priorities in the company's estimation of value can change over time. This would mean that the associated formulas for the risk analysis produce different values.

\subsubsection{Risk monitoring of the released application}
\label{sec:monitoring_relaesed}
%In addition to keeping the risk analysis up to date, it is also possible to try to monitor the usage of the specific instances of the protected application. To be able to perform this type of monitoring, an application needs to send information about its operation back to the vendor. This is of course only possible in the case where the application already had an online component to begin with, or if the deployment of a {\softprot} with an online component was deemed necessary by the risk analysis.

Next, one needs to track how copies of the released software are running on the premises of their users. 
This can be achieved by monitoring the information that the protected application communicates to the vendors. 
%This information can originate from the original application in case it already had an online component, as is the case in a streamed media client, or when a \softprot with an online component was injected as part of the mitigation process. 
Such information may originate from a monitoring-by-design \softprot such as reactive remote attestation~\cite{viticchie2016reactive}, but also from communication by other online components that were not originally designed for online monitoring.
This is particularly the case when irregular communication patterns can be linked to unauthorized activities such as running multiple copies in parallel or executing program fragments in a debugger in orders or frequencies not consistent with authorized uses. 
Such patterns can occur from communications present in the original applications, or from online {\softprot}s such as code renewability~\cite{renewability} and client-server code splitting~\cite{barrierslicing}. 
Importantly, the use of non-monitoring communication does not require the implementation of reaction mechanisms in the protected application to be effective. 
In many cases, it can suffice to feed the insights from the monitoring into the processes for keeping the risk-analysis up-to-date, or the application servers can take action.

%{\softprot}s with an online component might be designed explicitly to track how a program is used, but this needs not necessarily be the case in order to be applied to risk monitoring. An online component can serve a main purpose of protecting the application, where the online connection can be used to, for example, create a more powerful protection, or to allow for a fine-grained or application-specific reaction mechanism. The logs of the online components of such {\softprot}s can also be analysed to keep track of whether or not the application is being actively attacked, and by how many users.

%Furthermore, merely the way in which any online component interacts with the server for its primary function, can be analysed to create additional insights into how an application is (ab)used. For example, if a {\softprot}'s main functionality is to download code regions and execute them in the context of the application, the server can observe the \emph{order} in which a client downloads these code regions, and can thus also observe the code paths that were executed. These can then be analysed to check for inconsistencies and potential attacks. The logs of online components that were already part of the original, unprotected application, can similarly be re-purposed.

%The insights from this kind of monitoring do not necessarily need to be coupled to a specific reaction mechanism of the application. In this task, these insights serve to keep track of the exposition to risks of the protected software. This information can then be coupled back into the risk analysis.

Finally, the vendor of the released application needs to monitor whether the impacts of the deployed {\softprot}s  on the user experience and cost are in line with expectations or promises by the \softprot vendor. If, e.g., users start reporting usability issues, or if online protections manifest scalability issues, \eg, because more copies are sold than originally anticipated, those evolution might also warrant a revision of the risk mitigation strategy.

\if false
\section{Automation Requirements}

\changed{
It is on the mitigation task that automation poses the most severe constraints. 
First, optimizing the selection of the combination of {\softprot}s to apply in order to mitigate the risks must comply with strict computational constraints. In turn, this implies the solution of ad hoc optimization models that, given the very large size they may reach, must guarantee that results are obtained in useful time (minutes, hours, rarely days) and that pruning actually discards the less important combinations.

Moreover, an optimization process should decide based at least on the potency of the selected combination of protections and on the estimation of the performance of the protected app (\eg user experience). Current methods for estimating the potency and overheads are not usable for automatic decision support, as they require the actual application of the protections.
Given the time and resources needed to apply protections on non-toy programs, to measure objective metrics and compute or estimate the overheads by running the protected applications, the optimization would only consider a very limited solution space, making an optimization process useless.

Therefore, methods to predict the potency and the overheads are needed when combination of protections are applied.
As a solution, we have applied ML techniques to estimate the changes on the objective metrics after the application of protections. From the predicted metrics we compute the new potency. Also our simplified overhead estimation model is based on predicted metrics and data about the protections, nonetheless a new model that uses ML is under development.

As a last part that would complement an automatic decision support, we mention the automatic application of the selected protections. This step would avoid the complexity of learning how to properly configure the application of protections. Moreover, having such a framework would pave the road for an open standard for an API for software protection.
}

{\color{brown}

% For automating the risk management process, the ability is needed to automate the identification of:
\begin{itemize}

\item automatically identify the assets in the software is definitely outside the reach

\item 
identify the secondary assets to protect from an explicit specification of the primary assets in the software, \eg, by  using a sort of attacker model catalogue or as output of automatic attack discovery. 
Secondary target: show defenders that they protect much more than just assets, focus attention on a larger the important parts of the code, not just the assets

\item automatically determine which attack / attack steps can be automated and when manual effort is needed

\item ??? correctly model all the artifacts that tools produce (or may produce) to help attackers in mounting attacks. Associate some usability score. 

\item automatically identify the attacks/attack paths, the resulting findings must be rich enough to be usable for fine tuned decisions (\eg, CFF vs.\ OP). Includes: determining the attack strategies, adapt the automated discovery algorithms to the strategies.
Secondary target: teach people how hackers work, what they can do and why \softprot is important.
Secondary target: reassure experts that the engine works well
https://www.overleaf.com/project/5e1594c3b69e0300019fd9d9
\item 
automate the assessment of the attacks.\\
This requires the ability to estimate likelihood, complexity, (OWASP keywords). Other model of assessment (at least the fields) can be followed even if all the methods will be different.
Requires the ability to model what tools provide to attackers and when manual effort is needed.
May be hardcoded at the beginning, maybe more dynamic estimation can be achieved with more advanced automatic discovery algorithms

\item automatically identify the suitable protections based the attack assessment

\item prioritize assets and protections based on the attack assessment (may be superseded by full optimization)

\item automatically identify the protections in synergy and compose them properly to maximize protection benefits

\item 
overhead estimation: evaluate the impact of the application of {\softprot}s in terms of performance and user experience 

\item estimate the impact of the protections on assets/applications (protection level) without actually applying the protections.
Needed to evaluate several combinations/protections to choose the optimum.
Needed since assets and protections are interleaved, it is not feasible to find a general model of prediction that is independent on the application to protect (universal metric, in netsec you just evaluate the fw and a few functionalities based on the service you want to protect)

\item
optimize the selection of the {\softprot}s to apply in order to mitigate the risks. Constraints: satisfy usability of the applications and the investments required 
Depends on previous items (impact without application and overhead estimation)
Do it before the end of the world (computational issues as for many optimization problems)

\item automatically mitigate the risks introduced by the mitigate phase itself (\ie asset hiding)
\item automatically apply protections

\item the monitoring (as aspect of maintaining an up to date list of threats and risks and estimations and risk assessment procedures) depends on the economic model of the assets
\item 
\textcolor{red}{more?}
\end{itemize}
}
\fi

\section{Expert System for Software Protection}
\label{sec:workflow}
\label{sec:esp:kb}

The \esp is our \poc tool to implement an automated risk analysis of software protection\footnote{The full ESP code is available at \url{https://github.com/daniele-canavese/esp/}.}.
It is mostly implemented in Java and packaged as a set of Eclipse plug-ins with a customised UI. 
It protects software written in C. As it needs access to the application source code to work, the target users are software developers or protection consultants.

% \abnote{a bit of history, the fact that we have implemented different pieces of the puzzle}

\begin{figure}[t]
	\centering
    \begin{tikzpicture}[node distance=1.25em, every node/.style={minimum width=15em}]
    \tikzstyle{framing}=[draw=Gray, rounded corners, left color=DodgerBlue!25!White, right color=White, shading angle=210]
    \tikzstyle{assessment}=[draw=Gray, rounded corners, left color=IndianRed!25!White, right color=White, shading angle=210]
    \tikzstyle{mitigation}=[draw=Gray, rounded corners, left color=Goldenrod!25!White, right color=White, shading angle=210]
    \tikzstyle{edge}=[draw=Black!65, thick, -latex]
    \tikzstyle{biedge}=[draw=Black!65, thick, latex-latex, rounded corners]

    \node[draw=Tomato!75!Black, semithick, double, left color=White, right color=Tomato!50!White, shading angle=0, minimum width=20em] (sources) {unprotected application (source code)};
    
    \node[framing, below=of sources] (annotation) {asset annotation};
    \node[framing, below=of annotation] (analysis) {source code analysis};
    \begin{scope}[overlay]
        \node[fit=(annotation) (analysis), draw=DodgerBlue, very thick, densely dashed, rounded corners, label={[rotate=90, anchor=south]left:\textsc{\shortstack{risk\\framing}}}] (framing) {};
    \end{scope}

    \node[assessment, below=of analysis] (threats) {threat identification};
    \node[assessment, below=of threats] (risks) {risk evaluation and prioritization};
    \begin{scope}[overlay]
        \node[fit=(threats) (risks), draw=IndianRed, very thick, densely dashed, rounded corners, label={[rotate=90, anchor=south]left:\textsc{\shortstack{risk\\assessment}}}] (assessment) {};
    \end{scope}

    \node[mitigation, below=of risks] (decision) {decision making};
    \node[mitigation, below=of decision] (deployment) {deployment};
    \begin{scope}[overlay]
        \node[fit=(decision) (deployment), draw=Goldenrod, very thick, densely dashed, rounded corners, label={[rotate=90, anchor=south]left:\textsc{\shortstack{risk\\mitigation}}}] (mitigation) {};
    \end{scope}
    
    \node[below = of deployment, draw=GreenYellow!75!Black, semithick, double, left color=White, right color=GreenYellow!50!White, shading angle=180, minimum width=10em, xshift=-2.5em] (protected) {protected application};
    \node[below = of protected, draw=GreenYellow!75!Black, semithick, double, left color=White, right color=GreenYellow!50!White, shading angle=180, minimum width=10em, xshift=5em] (server) {server-side logic};
    \begin{scope}[overlay]
        \node[fit=(protected) (server), draw=GreenYellow!75!Black, very thick, densely dashed, rounded corners, label={[rotate=90, anchor=south]left:\textsc{\shortstack{risk\\monitoring}}}] (monitoring) {};
    \end{scope}

    \node[right=of assessment, draw=Gray, thick, cylinder, shape border rotate=90, aspect = 0.65, minimum height=3.5em, minimum width=3em, cylinder uses custom fill, cylinder body fill=DarkOrange!25!White, cylinder end fill=DarkOrange!5!White] (kb) {KB};
    
    \draw[edge] (sources) -- (annotation);
    \draw[edge] (annotation) -- (analysis);
    \draw[edge] (analysis) -- (threats);
    \draw[edge] (threats) -- (risks);
    \draw[edge] (risks) -- (decision);
    \draw[edge] (decision) -- (deployment);
    \draw[edge] ($(deployment.south west) + (1em, 0)$) -- ($(protected.north west) + (1em, 0)$);
    \draw[edge] ($(deployment.south east) + (-1em, 0)$) -- ($(server.north east) + (-1em, 0)$);
    \draw[biedge, dotted] (protected) -- (server);
    
    \draw[biedge] (kb.north) |- (framing.east);
    \draw[biedge] (kb.west) -- (assessment.east);
    \draw[biedge] (kb.south) |- (mitigation.east);
\end{tikzpicture}
	\caption{The \esp workflow.}
	\label{fig:workflow}
\end{figure}

% Also for the \esp, an \emph{asset} is a software component (\ie a variable or a code region) with some security requirement that must be safeguarded against attackers. % (\eg a cryptographic key must be kept confidential). 

% We will use the term \emph{solution} to denote an ordered sequence of protections (and their related configurations) that can be used to safeguard all the security properties of an application's assets. We are particularly interested in finding the \emph{golden solution}, that is the most secure solution that can be applied to protect a software application. Ultimately, the job of the \esp is to find the golden solution without any human intervention.

Figure~\ref{fig:workflow} depicts the high-level workflow, with four phases corresponding to the four phases described in Section~\ref{sec:requirements}, namely, risk framing, assessment, mitigation, and monitoring.

The \esp starts with the \emph{risk framing} phase, where all the information about the risk analysis to perform are collected and stored in a \kb. Context information includes general information, like the attacker model and the protections available to mitigate the risks, and information about the application to protect, like the assets and abstract representation of the application code, which is collected through code analyses. More details on the \esp risk framing are presented in Section~\ref{sec:esp:risk_framing}.

{Next}, the \esp performs the \emph{risk assessment} phase, whose details are provided in Section~\ref{sec:esp:risk_assessment}. 
It infers the attacks against the  assets and assesses the risks against each asset by estimating, for each inferred attack, the complexity to successfully execute it. 
The risk is evaluated by taking into account the structure of the application and the attacker model, \ie, the skills an attacker interested in endangering the application assets is likely to have, and asset values, as defined by the user during the risk framing. 
%The prioritization is made based on the asset values. %\bdsnote{What about the risk evaluation and prioritization? That part from the figure is not mentioned.} \lrnote{Added a short sentence to address this.}
   
The \esp's \emph{risk mitigation} phase, detailed in  Section~\ref{sec:esp:risk_mitigation}, uses \gls{ml} and optimization techniques to select the best \emph{solution}, i.e., the best sequence of {\softprot}s to be deployed and their configurations. It then automatically deploys it on the application code to generate the protected application binaries. 
The \esp can also be configured to propose a set of solutions that experts can manually edit to have full control over the protection deployment.
Moreover, the \esp can evaluate the effectiveness of solutions manually proposed by experts.
If remote protections are included in the selected solution, the deployment phase also generates the server-side logic, to be executed on a trusted remote entity.

Finally, the \emph{risk monitoring} is performed. However, the \esp does not dynamically update the risk analysis process parameters, it only performs real-time integrity checking (see Section~\ref{sec:monitoring_relaesed}).

\subsection{Risk framing}
\label{sec:esp:risk_framing}

%%%%%%%%%%%%%%%%%%%%%%%%%%%%%%%%%%%%%%%%%
The \emph{risk framing} starts with a preliminary preparation of a \kb with all the application-independent information, named \emph{generic a-priori information}, \ie the core concepts and the information that is not related to the specific application to be protected but is relevant for framing the risk analysis process.
A-priori information includes (but is not limited to) the assets types, the supported security requirements, all the known attack steps, and the available protections, and the necessary data to evaluate risks and mitigations, which have been already discussed in Section~\ref{sec:framing}.
Then, the \esp performs a \emph{source code analysis} that populates the \kb with \emph{a-priori analysis-specific information} using the C Development Toolkit of the Eclipse platform.. %which include application code artifacts, components, and structure, as well as the primary assets, previously annotated by the developers, and the identified secondary assets.
The source code analysis reports all the \emph{application parts}, which are the variables, functions and code regions therein. 
The analysis determines additional information, like variables' data types, function signatures, and produces additional representations, such as the call graph, which are useful to make decisions about the protections to apply.

The \poc of the \esp supports two requirements: confidentiality and integrity. 
The user needs to annotate the source code with pragma and attribute annotations to identify the assets in the code and to specify their security requirements formally. We designed a specific format for these annotations~\cite{D5.11,D5.13}.
The ESP then uses the call graph information to identify potential secondary assets.
In this phase, the \esp also puts the preferences and analysis customization parameters in the \kb before starting the rest of the process (\eg, the protections to consider, the selected attackers to counter).

To allow the automation of the risk analysis process, the \kb has been formally described by means of meta-model for software protection \cite{reganoMeta}, whose core classes are represented in Fig.~\ref{fig:metaModel}. The aspects corresponding to those classes will be discussed in the next sections. 
The \kb is instantiated as an OWL~2 ontology.
Together with a-priori information, there is \emph{a-posteriori information}, \ie, data inferred and stored during later work flow phases (\eg, the inferred attacks, the solutions).

\begin{figure}[t]
	\centering
    \begin{tikzpicture}[node distance=1em, every node/.style={minimum width=1em, minimum height=1em}]
    \tikzstyle{a-posteriori}=[draw=Gray, rounded corners=2, left color=YellowGreen!25!White, right color=White, shading angle=90]
    \tikzstyle{use-a-priori}=[double, draw=Gray, rounded corners=2, left color=Goldenrod!25!White, right color=White, shading angle=270]
    \tikzstyle{generic-a-priori}=[double, draw=Gray, rounded corners=2, left color=RedOrange!25!White, right color=White, shading angle=270]
    \tikzstyle{association}=[draw=Black!65, thick, -latex, rounded corners]
    \tikzstyle{composition}=[draw=Black!65, thick, -Diamond, rounded corners]
    \tikzstyle{generalization}=[draw=Black!65, thick, -{Triangle[open]}, rounded corners]
    
    \node[a-posteriori] (solution) {Solution};
    \node[a-posteriori, below=of solution] (apiItem) {AppliedProtectionInstanceItem};
    \node[a-posteriori, below=of apiItem] (api) {AppliedProtectionInstance};
    \node[use-a-priori, left=3em of api] (part) {ApplicationPart};
    \node[use-a-priori, above=of part] (app) {Application};
    \node[use-a-priori, below=of part] (asset) {Asset};
    \node[generic-a-priori, below=of asset] (requirement) {\shortstack{\emph{\footnotesize$\ll$enumeration$\gg$}\\SecurityRequirement}};
    \node[generic-a-priori, below=of api] (pi) {ProtectionInstance};
    \node[generic-a-priori, below=of pi] (protection) {Protection};
    \node[a-posteriori, below=of requirement, xshift=7.5em] (target) {AttackTarget};
    \node[a-posteriori, right=of target] (step) {AttackStep};
    \node[a-posteriori, below=of step, xshift=3.5em] (stepItem) {AttackStepItem};
    \node[a-posteriori, left=of stepItem] (path) {AttackPath};
    \node[generic-a-priori, below=of requirement, xshift=-3.75em, minimum width=1.25em, font=\footnotesize] (generic) {};
    \node[use-a-priori, below=0.25em of generic, minimum width=1.25em, font=\footnotesize] (use) {};
    \node[a-posteriori, below=0.25em of use, minimum width=1.25em, font=\footnotesize] (post) {};
    \node[right=0.075em of generic, minimum width=2em, font=\footnotesize] (generic2) {generic a-priori};
    \node[right=0.075em of use, minimum width=2em, font=\footnotesize] (use2) {a-priori use-case};
    \node[right=0.075em of post, minimum width=2em, font=\footnotesize] (post2) {a-posteriori};
    \node[draw=Silver, rounded corners, inner sep=1pt, fit=(post) (post2) (use) (use2) (generic) (generic2)] (legend) {};
    
    \draw[association] (solution) -- (apiItem);
    \draw[association] ($(apiItem.north) + (1.55, 0)$) -- ($(apiItem.north) + (1.55, 0.3)$) -- ($(apiItem.north) + (2.05, 0.3)$) -- ($(apiItem.north) + (2.05, 0)$);
    \draw[association] (apiItem) -- (api);
    \draw[association] (api) -- (part);
    \draw[composition] (part) -- (app);
    \draw[generalization] (asset) -- (part);
    \draw[association] (asset) -- (requirement);
    \draw[association] (api) -- (pi);
    \draw[association] (pi) -- (protection);
    \draw[association] (target) |- (asset);
    \draw[association] (target) -| ($(requirement.south) + (1.25, 0)$);
    \draw[association] (step) -- (target);
    \draw[association] (stepItem) |- (step);
    \draw[association] ($(stepItem.north) + (0.5, 0)$) -- ($(stepItem.north) + (0.5, 0.3)$) -- ($(stepItem.north) + (1, 0.3)$) -- ($(stepItem.north) + (1, 0)$);
    \draw[association] (path) -- (stepItem);
\end{tikzpicture}
	\caption{The \esp meta-model.}
	\label{fig:metaModel}
\end{figure}

{
% \color{green!50!black} 

In addition, the \esp offers a GUI to edit the framing information (\eg marking  additional assets, characterizing the attacker, and selecting protections) through a GUI, which also allows importing and exporting risk framing data as XML or OWL files. 
This feature was appreciated during the validation as it allows augmenting the analysis with important information that may be missed by the automatic process, like the secondary assets that might be linked into a protected program as part of certain protections.}

%During this phase, the \esp parses the source code for the annotations that mark the assets and their  security requirements in the source code. Annotations are source code constructs manually added by the developers via ad-hoc C pragmas and GCC-like variable attributes. \dcnote{Should we put an annotation example here? Not sure.}  \abnote{add cite for annotations}
%Moreover, primary and secondary assets can also be selected from the list of discovered application parts from an \esp GUI.  \bdsnote{And they can also be added to the knowledge base through JSON or XML files, not? I think this is important for when code injected by the tool chain after this phase needs to be protected as well.} \dcnote{Yes, you can! The knowledge base is an OWL2 ontology, which is an XML file. You can edit it as you wish, also using an OWL editor like protege (https://protege.stanford.edu).}

\subsection{Risk assessment}
\label{sec:esp:risk_assessment}
In the \emph{threat identification}, the \esp finds the attacks that can breach the primary assets' security requirements and stores them in the \kb. 
This stage is roughly equivalent to the ISO27k ``identify risk'' step (see Sections~\ref{sec:standardizedRisk} and~\ref{sec:assessment}). 

The identified attacks are represented as a set of \emph{attack paths}, \ie, ordered sequences of atomic attacker tasks called \emph{attack steps}. Attack paths are equivalent to attack graphs~\cite{attack_graphs} and can serve to simulate attacks with Petri Nets~\cite{petri_nets_attacks}. 
The attack steps that populate our \poc\ \kb originate both from a study and taxonomy by Ceccato\etal~\cite{ceccatoTaxonomy,emse2019} and data collected from industrial \softprot experts who participated in the European ASPIRE research project (\url{https://www.aspire-fp7.eu}).

The attack paths are built via backward chaining, as proposed in earlier work~\cite{basileOTP,reganoProlog}, and implemented with SWI-Prolog.
An attack step can be executed if its premises are satisfied, and produces conclusions, the results of the successful execution of that step.
The chaining starts with steps that allow reaching an attacker's final goal (\ie security requirement breach) and stops at steps without any premise.
As the search algorithm has exponential complexity ({it builds} a proof tree with increasing depth and width), the \esp implements basic {(i.e., aggressive and quite brutal)} mechanisms {to prune} the search space (\eg, maximum length of the inferred attack paths)~\cite{ReganoPhd}.

\if 0
An example of an attack path produced by the \esp is represented in Figure~\ref{fig:attackTree}.\bdsnote{If this reference of Fig 3 is not followed up by a discussion of what the reader should note in it or remember from it, then why do we include the figure and the reference at all? What value does it offer?}

\begin{figure}[h]
    \centering
    \begin{itemize}
        \footnotesize
        \item breach variable integrity
        \begin{itemize}
            \item modify the variable dynamically
            \begin{itemize}
                \item locate the variable using dynamic analysis
                \begin{itemize}
                    \item launch the application
                \end{itemize}
                \item locate the variable using static analysis
            \end{itemize}
            \item change statically the variable
            \begin{itemize}
                \item locate the variable using dynamic analysis
                \begin{itemize}
                    \item launch the application
                \end{itemize}
                \item modify the variable statically
            \end{itemize}
        \end{itemize}
    \end{itemize}
    \caption{An attack tree example. \dcnote{Replace the tree with a list}}
    \label{fig:attackTree}
\end{figure}
\fi

The \esp performs the \emph{risk evaluation and prioritization} by assigning a \emph{risk index} to each identified path. To that extent, every attack step in the \kb is associated to multiple attributes, including the \textit{complexity} to mount it, the minimum level of \textit{skill} required, the availability of support \textit{tools}, and their \textit{usability}.  Additional attributes can be associated to entities trivially. 
Each attribute assumes a numeric value in a five-valued range. 
For assessing the actual risks, the values of complexity metrics computed on the involved assets are used as modifiers on the attributes. 
For instance, an attack step considered of medium complexity can be downgraded to lower complexity if the asset to compromise is considered simple, e.g., because its cyclomatic complexity is below a custom threshold value.

The risk index of an attack path is obtained by aggregating the modified attributes of the composing attacks steps into a single value. Our \poc is rather simple: per attack step, it first aggregates all the step's modified attributes into a single attack step risk index; {then the attack path risk index is computed by multiplying the composing attack step indexes}. Other aggregation functions are supported, such as summing the steps' indexes, selecting maxima, and more complex features can easily be incorporated. One idea is to let the attack path risk index depend on how many different expert tools are required.

The attack path and the risk indices computed by our \poc\ \esp were welcomed by security experts (see Section~\ref{sec:esp:results}), amongst others because they served as an excellent starting point for evaluating the weaknesses of a target application before performing a more manual risk mitigation. Nonetheless, experts were interested in refining the identified, most risky attack paths into more concrete sequences of attack operations, and in some cases they would have manually updated the risk indices.
Indeed, in our \poc\ \kb, the attack steps are coarse-grained, such as ``locate the variable using dynamic analysis'' and ``modify the variable statically''.
This is a main limitation and, as discussed in Section~\ref{sec:identification_threats}, it is an open research question how much refinement is needed.

\subsection{Risk mitigation}
\label{sec:esp:risk_mitigation}

Before presenting the risk mitigation process as performed by the \esp, we {will} introduce more precise definitions. A \textit{protection} is a specific implementation of a \softprot technique by a specific \softprot tool.
For instance, control flow flattening~\cite{wangFlatteningTechReport} as applied by Diablo in the \actc (\url{https://github.com/aspire-fp7/actc}) and by Tigress (\url{https://tigress.wtf/}) are considered two distinct protections~\cite{diablo,tigress}.
A \textit{protection instance (PI)} is a concrete configuration of a protection. 
The \esp can use the PI to drive the \softprot tool to apply a protection on a chosen application part.
Multiple different PIs can be defined for the same protection, depending on the available parameters. 
%, but only the most significant ones are selected to limit the number of instances to consider during the optimization. For example, Diablo takes as input a parameter that specifies the percentage of the target code that must be obfuscated. Meaningful PIs are obtained using 10\%, 20\%, 50\% however, not 21\% and \22\%.
%    
An \textit{applied protection instance (applied PI)} is the association of a PI to an application part, which states that the PI has been selected to be applied on an application part. %\bdsnote{Includes configuration parameters?} %\abnote{I don't like DPI but...}\dcnote{In the original ADSS I called them APIs (applied PIs).}
A \textit{candidate solution} is then an ordered (because of compos\changed{a}bility and layering) sequence of applied PIs.   
%The order is introduced because application of protections is order-dependent. For instance, obfuscation can break the tamper-proof protections applied before it. %\bdsnote{the definition of PI already mentions that a PI is a set of parameters to deploy a protection on a chosen program part. Then what does the "association" mentioned here add to that? That is not clear to me.}
    % , specifying for each asset which protection techniques must be used to protect it, the protection tools that the \esp must drive to deploy such techniques, and the configuration parameters of the used tools.
%\end{itemize}

The \esp first searches for the \textit{suitable protections}, which are defined as the protections that impact any of the attributes of listed attack step, \eg, they are able to defer an attack step.
% A protection is deemeed suitable to defer an attack path if it is able to defer at least one attack step on the path.
Each PI is associated to a formula that alters these attributes for each attack step. 
Therefore, after the application of a protection, the risk index of the attack steps and paths are re-assessed. 

The formulas also consider a number of complexity metrics computed on the protected assets' code. This way, our approach incorporates Collberg's prescription of \emph{potency}~\cite{collberg1997taxonomy} as a measure of the additional effort that attackers will have to invest as they face protected, more complex code. 
%that are best able to defer the previously found attack paths.
The information about the impact of protections on attack steps, i.e., the parameters to be used in the formulas, are stored in the \kb. 
It is based on a survey among the developers of all protections integrated in the ASPIRE project~\cite{D5.11}, whom we asked to score the impact of their protections on a range of attack activities in terms of concrete, clearly differentiated forms of impacts. 
These include impact on human comprehension difficulty by increasing code complexity, impact by moving relevant code fragments from the client-side software to a secure server not under control of an attacker~\cite{viticchie2016reactive,renewability,codeMobility}, impact on difficulty of tampering through anti-tampering techniques with different reaction mechanisms and present monitoring capabilities~\cite{viticchie2016reactive}, and impact of preventive protections such as anti-debugging~\cite{diabloSelfDebugging,circulardebugging}. 
The survey results were complemented with security expert feedback, and validated in pen test experiments with professional and amateur pen testers \cite{ceccatoTaxonomy,emse2019}.

To model the impact of layered protections when recomputing the risk indices and of synergies between protections, additional modifiers are activated when specific combinations of PIs are applied on the same application part. The existence of synergies was also part of the mentioned survey. 

Candidate solutions must also meet constraints on the performance degradation and other forms of overheads. Our \poc filters candidate protections using five overhead criteria: client and server execution time overheads, client and server memory overheads, and network traffic overhead. 

Finally, the \textit{protection index} associated to a candidate solution is computed based on the recomputed risk indices of all the discovered attack paths against all application assets, weighted by the importance associated to each asset.

\subsubsection{Asset protection optimization approach}

The \esp then finds the mitigations by building an optimization model that is solved with a game-theoretic approach. The \esp tries to combine the suitable protections to build the optimal layered protection solutions, that is, it finds the candidate solution that maximizes the protection index and satisfies the constraints.

Computing the protection index by re-computing the risk index, requires knowledge of the metrics on the protected application. As applying all candidate solutions would consume an infeasible amount of resources, we have built a \gls{ml} model to estimate the metrics delta of applying specific solution without building the protected application~\cite{reganoMetric}. 
The model available in the \poc\ \esp has been demonstrated to be accurate for predicting variations of up to three PIs applied on a single application part. With more protections, however, the accuracy starts decreasing significantly, nonetheless, this issue seems to be solvable with larger data sets and more advanced \gls{ml} techniques.

The \esp uses the same predictors to estimate the overheads associated with candidate solutions. Per PI and kind of overhead, the \kb stores a formula for estimating the overhead based on complexity metrics computed on the vanilla application. These formulas were determined by the developers of the different protections integrated in the ASPIRE project.  

Having to deal with combinations greatly increases the solution space. To explore that space efficiently and to find (close to) optimal solutions in an acceptable time, the \esp uses a game-theoretic approach, simulating a non-interactive \softprot game. 
In the game, the defender makes one first move, \ie, proposes a candidate solution for the protection of all the assets. Each proposed solution yields a base protection index, which has a positive delta over the risk index of the unprotected application. 
The attacker then makes a series of moves. Each move corresponds to the investment of (some imaginary unit of) effort in one attack path, which the attacker selects from the paths found in the attack discovery phase. Similarly to how potency-related formulas of the applied protections yield a positive delta in protection index as discussed above, we use resilience-related formulas that estimate the extent to which invested attack efforts eat away parts of the protection potency, thus decreasing the protection index. These formulas are also based on security expert feedback. We refer to Regano's thesis for more details on this game-theoretic optimization approach that uses mini-max trees and a number of heuristics to the search space for the best candidate solutions in acceptable times and with acceptable outcomes \cite{ReganoPhd}, as will be evaluated below.

After solving the game, the \esp  shows the user the best protection solutions found during the optimization, i.e., the best first moves by the defender, from which the user can choose one, for which the \esp will then invoke the automated protection tools to apply the solution, as explained in Section~\ref{sec:esp:workflow:solDep}.

\subsubsection{Asset hiding}
\label{sec:esp:workflow:hiding}

As already discussed in Section~\ref{sec:framing}, protections are not completely stealthy because they leave fingerprints. 
%Software protections are never completely stealthy, they introduce \textit{protection fingerprints} on the protected code, \ie, peculiar %structures or run-time behaviours that an attacker can try to identify to find the assets. 
%For instance, code obfuscated with the control flow flattening features easily recognised control flow graphs. %An attacker, looking at the control flow graph of the application binary, may easily identify such areas as potential assets, and concentrate his analysis on them.

In a previous paper \cite{reganoL2P}, we have raised this problem and proposed a solution, based on the refinement of existing protection solutions with additional protections deployed also on non-asset code regions. Those lure the attacker into analyzing such regions in lieu of the assets' code, thus hiding the assets from plain sight. We have devised three asset hiding strategies. In \textit{fingerprint replication} protections already deployed on assets are also applied to non-sensitive application parts to replicate the fingerprints such that attackers analyse more application parts. With \textit{fingerprint enlargement}, we enlarge the assets' code regions to which the protections are deployed to include adjacent regions such that attackers need to process more code per protected region. With \textit{fingerprint shadowing}, additional protections are applied on assets to conceal the fingerprint of the originally chosen protections, to prevent leaking information on the security requirements.

The \poc \esp hides the protected assets as an additional decision making step. 
In this step, we add \textit{confusion indices} to the protection indices, which are computed by an ad hoc formula built to estimate the additional time needed by the attacker to find the assets in the application binary, after the application of hiding strategies. Also the computation of the confusion index requires the estimation of the complexity metrics of the application code after the application of the protections.
To build this model, we have studied the effects of the hiding strategies for the protections devised during the ASPIRE project. The results of this study, stored in the \esp \kb, are used to compute the confusion index.

Starting from the solutions generated via the game-theoretic approach, the \esp proposes additional application parts to protect by solving a Mixed Integer-Linear Programming (MILP) problem, expressed as an heavily customized instance of the well-known 0-1 knapsack problem \cite{knapsack-book} that maximizes the confusion index and uses overhead as weight in constraints.
%\bdsnote{I think we need a reference to a book or such on using MILP to solve the knapsack problem.}
%\abnote{a found a few, but given that it is quite well accepted that KP can be fomulated as an ILP, I would save that space}
%
The MILP problem is solved using an external solver, the \poc\ \esp supports lp\_solve\footnote{See \url{http://lpsolve.sourceforge.net/5.5/}.} and IBM CPLEX Optimizer\footnote{See \url{https://www.ibm.com/analytics/cplex-optimizer}.}.

\subsubsection{Deployment}
\label{sec:esp:workflow:solDep}

The final step in the \esp workflow is the application of the solution on the target application. The solution is chosen by the user amongst the ones presented by the \esp. The result of this step (and of the whole workflow) is the protected binary plus the necessary server side components for online protections, if any, ready to be distributed to final users. 
The \esp deploys a solution by driving automatic protection tools. At time of writing, the \esp supports Tigress, a source code obfuscator developed at the University of Arizona, and the \actc, which automates the deployment of protection techniques developed in the ASPIRE FP-7 project~\cite{D5.11,D5.13}. Table~\ref{tab:protections} summarizes the protection techniques supported by the \esp.

\begin{table}
    \centering
    \begin{tabular}{lcccc}
        \toprule
        \multirow{2}{*}{\textsc{protection type}} & \multicolumn{2}{c}{\textsc{requirements}} & \multicolumn{2}{c}{\textsc{tool}} \\ 
        \cmidrule(lr){2-3} \cmidrule(lr){4-5}
        & \textsc{conf.} & \textsc{integ.} & \textsc{ACTC} & \textsc{Tigress}\\
        \midrule
            anti-debugging             & \faCheckCircle & \faCheckCircle & \faCheckCircle & \faCircleO\\
            branch functions           & \faCheckCircle & \faCircleO     & \faCheckCircle & \faCircleO\\
            call stack checks          & \faCircleO     & \faCheckCircle & \faCheckCircle & \faCircleO\\
            code mobility              & \faCheckCircle & \faCheckCircle & \faCheckCircle & \faCircleO\\
            code virtualization        & \faCheckCircle & \faCheckCircle & \, \textcolor{gray}{\faCheckCircle}\tablefootnote{The ACTC provides limited support for code virtualization, meaning that it is not reliably applicable to all code fragments. Hence the ESP does not consider it a potential protection instance.} & \faCheckCircle\\
            control flow flattening    & \faCheckCircle & \faCircleO     & \faCheckCircle & \faCheckCircle\\
            data obfuscation           & \faCheckCircle & \faCircleO     & \faCheckCircle & \faCheckCircle\\
            opaque predicates          & \faCheckCircle & \faCircleO     & \faCheckCircle & \faCheckCircle\\
            remote attestation         & \faCircleO     & \faCheckCircle & \faCheckCircle & \faCircleO\\
            white-box crypto     & \faCheckCircle & \faCheckCircle & \faCheckCircle & \faCircleO\\
        \bottomrule
    \end{tabular}
    \caption{Protection techniques supported by the \esp, with enforced security requirements (Confidentiality and Integrity) and tools used to deploy the techniques. {For each tool, we only mark techniques supported on our target platforms, i.e., Android and Linux running on ARMv7 processors.}}
    \label{tab:protections}
\end{table}

{
%\color{green!50!black} 
Finally, we point out that the \esp has been engineered to be extensible. All the modules can be replaced with alternative components. For example, the risk assessment based on backward reasoning could be replaced with a more advanced attack discovery tool, the only constraint being {that it needs to produce} output {that is} compliant with the software protection meta-model. %\bdsnote{The above sentence needs to be completed}
Moreover, it is also possible to support new protections. 
It is enough to add all the required information into the \kb (\eg, evaluation of strengths and impacts on attack steps, conflicts and synergies with other protections plus all parameters of the discussed formula). The only demanding operation is training the \gls{ml} algorithms to predict how new protections alter the metrics, and the automation of the deployment of the protections.

%simply adding a new \esp module that correctly implements the API calls needed to drive the protection tools, \eg stating the correct way to instruct the tool to protect a specific function with a specific protection instance exposed by the tool, and adding in the Knowledge Base the information related to the protection techniques supported by the tool, \ie the related protection instances. Particular care should be exercised in evaluating the compatibility among new protections and the existing ones. The \esp Knowledge Base permits to express precedence relationships and incompatibilities amongst protection instances. For example, the \esp has been instructed trough precedence relationships to first deploy source code obfuscations with Tigress, since the ACTC includes supports binary obfuscation techniques (through the Diablo obfuscator by Gent University) that must be obviously deployed afterwards.
}

\subsection{Risk monitoring}
\label{sec:esp:risk_monitoring}

% \abnote{for symmetry purposes we may have to resort to this section}
The \esp generates all the server-side logic, which also includes the backend\changed{s} necessary to the protections that are used to perform the risk monitoring of the released application.

Our \poc\ \esp does not automatically include the feedback and other monitoring data (\eg number and frequency of detected attacks and compromised applications, server-side performance issues)
The knowledge base needs to be manually updated by the experts, even if GUIs can be used to save time, to change risk framing data related to attack exposure and protection effectiveness. Also, issues related to insufficient server resources need to be addressed independently, the \esp only provides the logic, not the server configurations.

\subsection{Validation}
\label{sec:esp:results}

The \esp has been validated with experts, mainly from the ASPIRE project consortium and advisory boards, to assess the quality of the identified solutions and fulfillment of the requirements for the software risk management methodologies, as in Section~\ref{sec:requirements}. Moreover, we have evaluated its performance to determine computational feasibility and scalability of its tasks.  

\subsubsection{Qualitative evaluation}
As a first qualitative evaluation, software security experts tasked with the protection of a target application have been asked to judge the correctness of the workflow, suitability with their daily software protection workflow and tasks, potential impact on the quality of their work, and effectiveness of their results.
Overall, experts have judged the \esp as very promising and potentially effective by the experts to delegate or support their tasks, because of the high-level of automation and configurability, of the possibility to override default configurations, of the very detailed output,
Nonetheless, they were skeptic on the usability by software developers with a limited background on software protection.
Among the data extracted by the tool, experts highlighted the importance of making decisions by taking into account the application structure and metrics, because results are tailored to the target application.
They also appreciated that all the data extracted and represented in the \kb are structured according to a meta-model.

{During the ASPIRE project, each industrial partner provided an Android application, namely a One-Time Password generator for home banking applications, an application licensing scheme, and a video player for \drm protected content. Each Android application included security-sensitive code in a dynamically linked library written in C. It were these libraries that served as reference use cases for all the protection tasks.}
In a qualitative evaluation, we asked two experts from each of the three industrial project partners to validate the solution selected by the \esp for their application.  %have been involved in a more in-depth analysis of the \esp in action. 
%The experiment consisted in asking the 
{While} details {of} these use cases, which were designed to embed relevant assets but without having the full complexity of real-world applications, have not been publicly released by their proprietors, they are included in a confidential project deliverable~\cite{D1.06.9}. Table~\ref{tab:useCase} discloses the \sloc metrics {and the number of assets} as evidence that the \esp was not used merely on toy examples. 

% The data produced by \esp analysed by the 
Experts analysed the inferred attack paths and protections to mitigate them, and the solutions proposed by the optimization process, {comparing them to solutions they manually assembled earlier on during the project as part of the requirements formulation}.
Solutions have been validated in terms of achieved security for the assets, preservation of the application business logic, and containment of the inevitable slow-down of the protected application \wrt the original one. 
Furthermore, the attack paths have been compared with the real attacks discovered by the professional tiger teams previously involved in the assessment of the protections.

Also this validation gave a positive output: quoting from the related project deliverable, \textit{``after the analysis of the validation data, the experts concluded that the tool has a very high potential''} even if some had doubts on the maturity of the tool to use it, in its current form, on their existing commercial applications with all of their SDLC intricacies and complexity. {For a research proof-of-concept, this should of course not come as a surprise.}
In particular, the inferred solutions have been judged as appropriate to protect the use case code and effective to block the inferred attack paths and the real attacks reported by the tiger teams. Moreover, the protected binaries were evaluated as being semantically unaltered (\eg, still delivering the original observable IO-relation) and usable (\ie, without an excessive overhead introduced by the protections).
The main flaw of \esp reported by the experts is that inferred attack paths were too coarse-grained, because of too generic attack rules.  %future work on \esp will have to address this limitation, expanding the existing attack rules, \eg by modelling actual attacks against real application listed in databases such as the CVE\footnote{\url{https://cve.mitre.org/}} and CWE\footnote{\url{https://cwe.mitre.org/}}, or in automated attack frameworks such as Metasploit\footnote{\url{https://www.metasploit.com/}}. The complete report on the \esp validation is presented in the related project deliverable \cite{D1.06.9}.

%,  All the concepts and relationships between them that are deemed useful in the software protection process have been formalized in a software security meta-model, presented in Section~\ref{sec:esp:kb}. Also, protection solutions are inferred taking into account the complexity metrics computed on the code comprising each asset, thus the efficiency of the solutions in safeguarding the assets reflects the real characteristics of the code.

% Also, the implemented  to the \esp user. In particular, the latter needs to provide to \esp the assets and the security requirements for them, which h. Also, \esp can deploy the inferred protection solution automatically, since all the evaluated protections are implemented with automatic protection tools (Diablo, Tigress, ACTC). Thus, \esp can be used also by software developers with a limited background in software security. Also, the results of each phase in the \esp workflow are provided to the user in a human-readable format, \eg sequences of simple attacker tasks for the risk assessment phase, ordered lists of protection techniques to be applied for each asset in the risk management phase. In this way, an expert using \esp can obtain automatically a protection solution for the application that he must protect, and can then evaluate it analyzing the results produced in each \esp workflow phase. If needed, he or she can then refine the automatically obtained solution.

\begin{table*}[t]
	\centering
	\begin{tabular}{lrrrrrr}
		\toprule
		\multirow{2}{*}{\textsc{application}} & \multicolumn{2}{c}{\textsc{C}} & \multirow{2}{*}{\textsc{Java}} & \multirow{2}{*}{\textsc{C++}} & \multirow{2}{*}{\textsc{assets}}\\
		\cmidrule(lr){2-3}
		& \textsc{sources} & \textsc{headers} & & &\\
		\midrule
		DemoPlayer & 2,595 & 644 & 1,859 & 1,389 & 25 \\
		LicenseManager & 53,065 & 6,748 & 819 & 0 & 43 \\
		OTP & 284,319 & 44,152 & 7,892 & 2,694 &  25 \\
		\bottomrule
	\end{tabular}
	\caption{Lines of source code counts of the ASPIRE validation use cases.}
	\label{tab:useCase}
\end{table*}

\subsubsection{Experimental assessment}

%\abnote{Substitute A,B,C with the real app}

{
We have measured the execution time of the \esp on the three use cases, with the assets annotated by the experts, as reported in Table~\ref{tab:useCase}. In all the three cases, 17 protection instances have been considered. These are the nine supported protections of the ACTC as listed in Table~\ref{tab:protections}, of which opaque predicates, branch functions and control flow flattening can be applied at three configuration levels (i.e., low, medium, and high frequencies with corresponding high levels of overhead), and in which three different data obfuscations were considered (XOR-masking, residue number encoding, and data-to-procedural conversions) \cite{collberg1997taxonomy}.

Table~\ref{tab:adss-times} shows the \esp computation times. The framing phase is almost instantaneous and is driven by the lines of the code from which annotations are extracted. Regarding the complexity and scalability of the assessment and the mitigation phases, it is not possible to draw conclusions from this experiment use cases that were driven by the industrial partners, aimed at testing the \esp in a real scenario, and of which we did not control the variables that we deemed interesting for a complete performance evaluation.% have mixed results. The \sloc and the number of assets .

%\textcolor{red}{MORE}.

\begin{table}[t]
	\centering
	\begin{tabular}{lcccc}
		\toprule
		\textsc{application} & \textsc{Total} & \textsc{Framing} & \textsc{Assessment} & \textsc{Mitigation} \\
		\midrule
		DemoPlayer & 145.6 & 0.1 & 76.3 & 69.1\\
		LicenseManager & 296.1 & 0.3 & 187.6 & 108.1\\
		OTP & 170.0 & 0.94 & 69.2 & 99.9\\
		\bottomrule
	\end{tabular}
	\caption{ESP times in seconds.}
	\label{tab:adss-times}
\end{table}
}

{
Regarding the attack discovery tool and the game theoretic optimization of the mitigation phase, we know that the most influential factors of the assessment phase are the number of attack steps in the \kb (exponential complexity but with pruning) and the number of assets (linear), while the mitigation phase is driven by the number of assets (linear), the number of PIs and {and the number of attacks discovered in the assessment phase} (both exponential).

To asses the scalability of those phases, we evaluated the performance of the \esp on three standalone Linux applications that cover an increasing range of assets. Table~\ref{tab:expUseCase} summarizes their metrics.
These toy applications have been randomly generated with a process that selects a call graph (from a set of call graphs extracted from real applications), then generates randomized function bodies to meet some code metrics, and then randomly selects fragments in the generated code as data or code assets.
In this experiment, we have used all the previously listed PIs from the ACTC (excluding white-box crypto, which was a proprietary algorithm of one industrial ASPIRE project partner) and added four instances of obfuscation applied using Tigress, i.e., the ones marked in Table~\ref{tab:protections}.
}

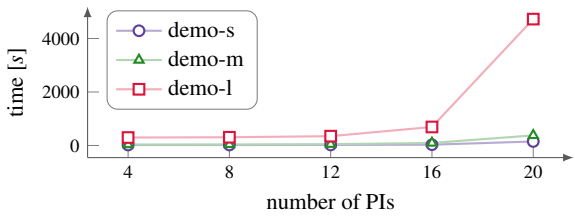
\begin{figure}[t]
%	\hspace*{-0.5in}
	\centering
	\begin{tikzpicture}
	\begin{axis}
	[legend style={at={(0.35,0.975)}, legend cell align={left}, anchor=north east, draw=gray, rounded corners}, height=12em, width=0.95\linewidth, axis y line*=left, axis x line*=bottom, axis line style={Black!65, -latex}, xlabel={number of PIs}, ylabel={time [$s$]}, xtick={4,8,12,16,20}]
	\addplot[RoyalPurple!35, thick, mark = *, mark options={solid, RoyalPurple, fill=white}, thick] coordinates {(4,17.399) (8,18.129) (12,20.27) (16,29.674) (20,149.757)};
	\addplot[ForestGreen!35, thick, mark = triangle*, mark options={solid, ForestGreen, fill=white}, thick] coordinates {(4,35.489) (8,38.726) (12,48.274) (16,93.432) (20,377.8)};
	\addplot[Crimson!35, thick, mark = square*, mark options={solid, Crimson, fill=white}, thick] coordinates {(4,297.485) (8,307.439) (12,348.038) (16,691.854) (20,4724.334)};
	\legend{demo-s, demo-m, demo-l} 
	\end{axis}
	\end{tikzpicture}
	\caption{\esp execution times on applications reported in Table~\ref{tab:expUseCase}.}
	\label{fig:espTime}
\end{figure}

On the three demo applications, the asset mitigation phase was repeated multiple times (see Section~\ref{sec:esp:risk_mitigation}), depending on the number of PIs available to protect the applications' assets. 
All experiments have been executed on an Intel i7-8750H 2.20~\si{GHz} computer with 32~\si{GB} RAM, using Java 1.8.0\_212 under GNU/Linux Debian 4.18.0. 
Figure~\ref{fig:espTime} depicts the measured total \esp computation time, along with the time needed for the risk assessment, asset protection, and asset hiding phases.
The time needed to complete the workflow increases with the number of PIs considered during the mitigation; such increase strongly depends on on the application code complexity (\eg \sloc, number of assets, and functions).

The time needed to analyze the applications source code and to generate the application meta-model instance was negligible (less than 1\si{s}), and  the time to deploy the solution is irrelevant for the assessment of {the} \esp' computational feasibility, as it only measures the time needed to execute the external protection tools on the single selected solution.

As expected, the time needed to execute the risk assessment phase does not depend on the number of PIs available to protect the application, as attacks are determined on the vanilla application. Nonetheless, we report that it has limited impact because of the aggressive pruning we have implemented that avoids the exponential growth.
The asset protection phase is by far the most computationally intensive, especially when the available PIs increase. 
Since, the mitigation considers sequences of protections, the execution times exponentially depends on the permutations of combination of PIs.
The same holds also for the asset hiding phase, even if less time needed to execute the latter, compared to the asset protection phase.

% \abnote{the reduction factor of the mitigations also summarizes the impact of the attacks that circumvent, undo, bypass the technique.}

\begin{table}[t]
	\centering
	\begin{tabular}{lrrrrr}
		\toprule
		\multirow{2}{*}{\textsc{Application}} & \multirow{2}{*}{\textsc{\sloc}} & \multirow{2}{*}{\textsc{functions}} & \multicolumn{3}{c}{\textsc{assets}}\\
		\cmidrule(lr){4-6}
		&  &  & \textsc{code} & \textsc{data} & \textsc{total}\\
		\midrule
		demo-s & 443 & 18 & 2 & 2 & 4 \\
		demo-m & 1029 & 47 & 12 & 3 & 15 \\
		demo-l & 3749 & 178 & 26 & 13 & 39 \\
		\bottomrule
	\end{tabular}
	\caption{Statistics of \acrshort{esp} experimental assessment applications.}
	\label{tab:expUseCase}
\end{table}

% \section{Daniele, aka Contribution on L1P}
% \label{sec:l1p}
% \input{l1p}

% \section{Bart aka Discussion }

% \section{Paolos} Unfortunately commented.

\section{Related work}
\label{sec:related}

Expert systems have been applied in the cybersecurity field since 1986, when Hoffman proposed an expert system for the risk analysis of computer networks \cite{hoffman1986risk}. The author theorized a system able to identify vulnerabilities in the analyzed computer system configuration and suggest the appropriate countermeasures to reduce the overall risk. %He stated that, even if in the early stage of development, using an expert system may lead to a positive outcome, especially after building a suitable general model for cybersecurity experts knowledge.
In the same year, Denning and Neumann started the development of IDES \cite{idesReq}, a host-based \ids mixing an expert system with statistical anomaly detection techniques aimed at detecting unauthorized accesses, both by local and remote users. %, of resources hosted by the monitored system. 
%A first prototype, monitoring a \acrshort{dec}\text{-2065} computer running the TOPS-20 operating system, has been presented in 1990 \cite{ides}. 
Its evolution, called NIDES \cite{nidesReq}, supported also real-time analysis of inter-process communications.

Other \ids expert systems were developed in the same years for specific tasks: NIDX \cite{nidx}, to suggest a network administrator possible security breaches of UNIX System V machines, NADIR \cite{nadir}, to monitor the internal network of the US Los Alamos National Laboratory, AUDES \cite{audes}, to assist computer security auditing process, and Haystack \cite{haystack}, to detect breaches in US Air Force systems. Notably, the latter was the first showing self-learning capabilities, as it evolved user profiles over time.
%AUDES \cite{audes} was developed to assist computer security auditing process, such as the analysis of log-in or resource access records after a security incident.

%Other \ids expert system were developed in the same years.  The \nidx \cite{nidx} was designed to suggest a network administrator possible security breaches of UNIX System V machines. Its knowledge base contained information about the structure of the monitored network, and usage profiles of its users, and was able to mimic the decision process of administrators. 
%The \nadir \cite{nadir} was built to monitor the internal network of the US Los Alamos National Laboratory, and Haystack \cite{haystack}, to detect breaches in US Air Force systems, the latter being the first showing self-learning capabilities, as it evolved user profiles over time. 
%The \audes \cite{audes} was developed to assist computer security auditing process, such as the analysis of log-in or resource access records after a security incident.

After this initial enthusiasm, the limitation of those expert systems in accuracy, manageability and extensibility, encouraged researchers to investigate combinations of classic expert systems with other techniques, in order to enhance breach detection performances. Examples are \cite{fwExpSys} by Eronen and Zitting, using constraint logic programming \cite{clp} to generate access control lists for Cisco firewalls from high-level filtering requirements, \cite{deprenIDS} by Depren\etal, using \som and decision trees for breach detection and an expert system to interpret the result of such machine learning algorithms, \cite{panIDS} by Pan\etal, employing neural networks for detecting attacks leveraging unknown vulnerabilities, while an expert system identifies known attacks.
%
%As an instance, Eronen and Zitting \cite{fwExpSys} presented a tool
%to analyze firewall rules which 
%using constraint logic programming \cite{clp} %to model high-level filtering requirements and translating it into a Cisco low-level access control list.
%to generate access control lists for Cisco firewalls from high-level filtering requirements.
%\footnote{\url{https://www.cisco.com/c/en/us/support/docs/security/ios-firewall/23602-confaccesslists.html}}. 
%Also, combinations of machine learning approaches and expert systems have presented. Examples are \cite{deprenIDS} by Depren\etal, using \som and decision trees for breach detection and an expert system to interpret the result of such machine learning algorithms, and \cite{panIDS} by Pan\etal, which uses neural networks for detecting attacks leveraging unknown vulnerabilities, while an expert system identifies known attacks.
Fuzzy logic algorithms have also been embedded in expert system for real-time intrusion detection, as in \cite{owensIDS}, and post-incident network forensics, for example in \cite{kimForensics} and \cite{liaoForensics}.
The EC-funded PoSecCo project has delivered SDSS \cite{posecco}, an expert system aimed at driving security administrator during all the steps from the policy specification,  anomaly analysis and resolution, to the automated refinement and enforcement of the anomaly-free policy. {Automated attack graph generation for microservice architectures that depend on potentially vulnerable third-party components have been proposed as well~\cite{10.1145/3297280.3297401}.}

{In the domain of \mate \softprot, there is a limited body of scientific literature on expert systems.} In practice, companies provide so-called cookbooks with protection recipes. For each asset, users of their tools are advised to deploy the relevant SPs in an iterative, layered fashion starting as long as the overhead budget allows for additional protections. Automated approaches are limited to specific types of protections, and hence only support specific security requirements. 
{Collberg, Thomborson, and Low~\cite{collberg1997taxonomy} and Heffner and Collberg~\cite{heffner2004obfuscation} studied how to overcome the problem of deciding which obfuscations to deploy in which order and on which fragments given an overhead budget. So did Liu et al.~\cite{obf_optvialangmods,7985664}. Their approaches are limited to obfuscation. They differ in their decision logic and in the metrics they use to measure protection effectiveness. Importantly, however, their used metrics are fixed and limited to specific program complexity and program obscurity metrics, without taking into consideration or adapting the used metrics to concrete potential attack paths.}
Coppens et al.\ proposed an iterative software diversification approach to counter a concrete form of attack, namely diffing attacks that exploit security patch releases to automatically engineer attacks against unpatched systems~\cite{coppens2013feedback}. In all of the mentioned works, obfuscations are the only considered {\softprot}s. In all works, measurements are performed after each round of transformations, much like in the second approach we discussed in Section~\ref{sec:decision_making}.
 
{To improve the user-friendliness of manually deployed \softprot tools, i.e., to make up to some extent for the lack of expert systems to select the {\softprot}s automatically, Brunet et al.\ proposed composable compiler passes and reporting of deployed transformations~\cite{10.1145/3338503.3357722}. Holder et al.\ evaluated which combinations of obfuscating transformations, and which order in which to apply them, yield the most effective overall obfuscation~\cite{obf_evaloptphaseord}. However, they did not discuss the automation of the selection and ordering given a concrete program with concrete security requirements.}

\section{Conclusion and future work}
\label{sec:conclusion}
We presented the necessity of having a standardized approach for risk management in the context of software protection against man-at-the-end attacks. To that end, we discussed just such a risk management approach for software protections, which we based on the NIST SP800-39 standard for risk management for information security. We discussed in detail how the different aspects of software protection can and should be mapped onto risk framing, risk assessment, risk mitigation, and risk monitoring phases. A proof-of-concept decision support system we have designed and implemented provides evidence that the proposed approach is feasible and can be automated to a large degree. We hope this is sufficient evidence that it is useful to launch a community effort that leads to a future standardization.

\bibliographystyle{spmpsci}
\bibliography{biblio}


\end{document}